\documentclass[iop]{emulateapj}

\usepackage{graphicx}
\usepackage{color}

\shorttitle{Circumstellar Light Echo as a Possible Origin of the Polarization of Type IIP Supernovae}
\shortauthors{Nagao et al.}

\begin{document}

\title{Circumstellar Light Echo as a Possible Origin of the Polarization of Type IIP Supernovae}

\author{Takashi Nagao\altaffilmark{1,3,4}, Keiichi Maeda\altaffilmark{1}, Masaomi Tanaka\altaffilmark{2}
}
\altaffiltext{1}{Depertment of astronomy, Kyoto University, Kitashirakawa-Oiwake-cho,
 Sakyo-ku, Kyoto 606-8502, Japan}
\altaffiltext{2}{National Astronomical Observatory of Japan, Mitaka, Tokyo 181-8588, Japan}
\altaffiltext{3}{Email: nagao@kusastro.kyoto-u.ac.jp}
\altaffiltext{4}{Research Fellow of Japan Society for the Promotion of Science (DC2)}


\begin{abstract}
Type IIP supernovae (SNe IIP) are the most common class of core-collapse SNe. They often show rapid increase of polarization degree in the late phase. This time evolution is generally believed to originate from the emergence of an inner aspherical core, while an effect of polarized-scattered echoes by circumstellar (CS) dust around the SN may also substantially contribute to this polarization feature. In this study, we examine the effects of the scatted echoes on the SN polarization through radiative transfer simulations for various geometry and amount of CS dust. It has been found that asymmetrically-distributed CS dust, which is generally inferred for red supergiants, could reproduce the observed polarization features. We have applied our results to SNe 2004dj and 2006ov, deriving the geometry and amount of CS dust to explain their observed polarization features in this scenario. For both SNe, the blob-like or bipolar distribution of CS dust rather than the disk-like distribution is favored. The derived dust mass $M_{\mathrm{dust}}$ in the blob model (the bipolar CS dust model) for SNe 2004dj and 2006ov are $\sim 7.5 \times 10^{-4}$ M$_{\odot}$ ($\sim 8.5 \times10^{-4}$ M$_{\odot}$) and $\sim 5.2 \times 10^{-4}$ M$_{\odot}$ ($\sim 1.3 \times10^{-3}$ M$_{\odot}$), respectively. Even in the case where this process would not play a dominant role in the observed polarization signals, this effect should in principle contribute to it, the strength of which depends on the nature of CS dust. Therefore, this effect must be taken into account in discussing multi-dimensional structure of an SN explosion through polarimetric observations. 
\end{abstract} 

\keywords{circumstellar matter - polarization - dust, extinction - radiative transfer - stars: mass-loss - supernovae: general}

\section{Introduction}
Core-collapse supernovae (SNe), which originate from death of massive stars ($\gtrsim 8 M_{\odot}$), play important roles in the Universe. They chemically pollute the interstellar space by ejecting a large amount of heavy elements, and also affect star formation via their huge energy output. However, the explosion mechanism is still unclear. It is now widely accepted that one-dimensional simulations could not reproduce the SN explosions \citep{Rampp2000, Liebendorfer2001, Thompson2003, Sumiyoshi2005}. Recently, successful explosions for some progenitor models have been reported by multi-dimensional simulations \citep[e.g.,][]{Buras2006, Marek2009, Suwa2010, Muller2012, Takiwaki2012, Hanke2013, Bruenn2013, Takiwaki2014, Couch2014, Melson2015, Lentz2015, Roberts2016, Muller2016}, even though the resulting explosion energy is still lower than the value typically inferred from observations ($\sim 10^{51}$ erg).  Multi-dimensional effect is believed to be a key to understanding the SN explosion mechanism \citep[e.g.,][]{Maeda2008, Tanaka2017}.

Polarimetric observations have been used to constrain the multi-dimensional geometry of SN explosions \citep[see][for a review]{Wang2008}. If a photosphere of an SN and the layer above it are spherically symmetric, polarization vectors due to electron scattering are completely cancelled out. Some continuum polarization would be observed when the photosphere deviates from spherical symmetry \citep[e.g.,][]{Shapiro1982, Hoflich1991, Hoflich1996, Kasen2006, Dessart2011, Bulla2015}. Type IIP SNe (SNe IIP) belong to the most common class of core-collapse SNe \citep[$\sim 50$ \% of the all core-collapse SNe in a volume-limited sample, e.g.,][]{Li2011}. Their progenitors are red supergiants (RSGs) with a thick hydrogen envelop. An optical light curve of an SN IIP shows a rapid rise in the early phase (the cooling phase), then a constant luminosity until $\sim 80$ days since the explosion (the plateau phase), followed by an exponential decline (the nebular phase) after a sudden drop \citep[e.g.,][]{Anderson2014, Sanders2015}. SNe IIP often show rapid increase in a continuum polarization level ($\sim 1$\%) just after entering into the nebular phase, following generally a small polarization level ($\sim 0.1$\%) in the plateau phase \citep[e.g.,][]{Leonard2006, Chornock2010, Kumar2016}. This polarimetric behavior could be explained as a highly asymmetric core revealed in the nebular phase. This is regarded as a supporting evidence for the asymmetric nature in the SN explosions as suggested by the above mentioned recent simulations.

There is, however, a possible alternative scenario to explain the polarization feature in SNe IIP. Scattering of SN light by aspherically-distributed circumstellar (CS) dust is another possibility to produce the net continuum polarization in SNe IIP \citep[hereafter dust scattering model; e.g.,][]{Wang1996, Mauerhan2017}. The polarized-scattered light by CS dust reaches to an observer with time delay. Thus, when the SN light suddenly becomes fainter, the relative contribution of the scattered light in the observed SN light becomes lager. \citet{Wang1996} have shown an excellent match between the dust scattering model and observed-temporal evolution of polarization in SN 1987A. Moreover, recent high-resolution observations of RSGs have revealed asymmetric and/or clumpy CS environment. High-resolution imaging and polarimetric observations of the most famous RSGs, Betelgeuse and Antares, indicate highly inhomogeneous CS environment \citep{Cruzalebes1998, Hinz1998, Marsh2001, Kervella2009, Kervella2011, Kervella2016, Ohnaka2009, Ohnaka2011, Ohnaka2013, Ohnaka2014}. There are similar reports for other RSGs: WOH G64, VY CMa and NML Cyg \citep[e.g.,][]{Wittkowski1998, Kastner1998, Smith2001, Monnier2004, Humphreys2007}.

In this study, we calculate time evolution of polarization through CS dust surrounding SNe IIP, discussing the possibility to explain the observed polarization feature for SNe IIP. Since the polarization feature is widely used to extract asymemtric nature of SN ejecta, it is important to clarify the effect and possible contribution of the CS dust on the observed feature to understand the explosion mechanism of SNe IIP. In Section 2, we describe our methods. In Section 3, we summarize our general results and apply them to SNe 2004dj and 2006ov. In Section 4, discussions are given for further details on the dust scattering models, inferred CS environment, and further prospects. Finally, conclusions are given in Section 5.

\section{Methods}
\subsection{Radiative Transfer}
We perform three-dimensional Monte Carlo radiative transfer calculations to study polarization due to scattering of SN light by CS dust. The simulation code, taking into account absorption and scattering by dust, is an update of the one presented in \citet{Nagao2016} with an additional capability of treating the polarization. This time-dependent transfer code applies to an arbitrary spatial distribution of the CS dust.

We use $100 \times 100 \times 100$ Cartesian meshes. For calculating polarization due to dust scattering, we use a widely-adopted method, following \citet{Code1995}. The Stokes vector ${\bf S}$ consists of the four Stokes parameters $(I,Q,U,V)$, where $I$ is the intensity, $Q$ and $U$ are the components of the linear polarization, and $V$ is the circular polarization \citep[see][]{Chandrasekhar1960, van de Hulst1957}. The polarization degree is given by $P=(Q^2 + U^2 + V^2)^{1/2}$, and the orientation of the linear polarization vector is given by the position angle $\chi = 0.5 \tan^{-1} (U/Q)$. We do not consider circular polarization $V$, which is expected to be negligible in the situation considered in this study \citep[e.g.,][]{Code1995, Patat2005}. As for the scattering matrix ${\bf A}$, we adopt the following approximation \citep{White1979}:
\begin{eqnarray}
{\bf A} = \frac{3}{4} \left(
    \begin{array}{ccc}
	P_1 & P_2 & 0\\
	P_2 & P_1 & 0\\
	0 & 0 & P_3
  \end{array}
  \right)
\end{eqnarray}
where,
\begin{eqnarray}
  \left\{
  \renewcommand{\arraystretch}{1.4}
  \begin{array}{l}
    \scalebox{1.0}{$\displaystyle
      P_1 = \frac{1-g^2}{(1+ g^2 - 2g \cos \theta)^{3/2}} $}\\
     \scalebox{1.0}{$\displaystyle
       P_2 = -p_l P_1 \frac{1- \cos^2 \theta}{1+ \cos^2 \theta} $}\\
     \scalebox{1.0}{$\displaystyle
	P_3 = P_1 \frac{2\cos \theta}{1 + \cos^2 \theta} $}.
   \end{array}
  \right.
\end{eqnarray}
Here, $g$ is a scattering asymmetry parameter, ranging from $0$ for isotropic scattering to $1$ for perfect-forward scattering, and $p_{l}$ is the peak linear polarization. Our code was tested with the results by \citet{Code1995}. We calculated dust scattering processes with the same dust parameters ($\omega \sim 0.5$, $g \sim 0.6$ and $p_{l} \sim 0.5$) in the situation that input parallel light is scattered by dust particles uniformly distributed in a spherical blob. The calculated flux and polarization degree of the scattered light as a function of scattering angle resulted in a good agreement \citep[Figure 5 in][]{Code1995}. In this study, we set the albedo $\omega = 0.5, g=0.6$ and $p_l = 0.5$ as the typical values for various dust models in the optical wavelength \citep[e.g.,][for a review]{Draine2011}, although the values are observationally uncertain.

\subsection{Distribution of CS dust}
For distribution of CS dust, three configurations are considered: the blob, disk and bipolar CS dust models (see Figure 1).

\begin{figure*}[t]
\begin{center}
\includegraphics[scale=0.2]{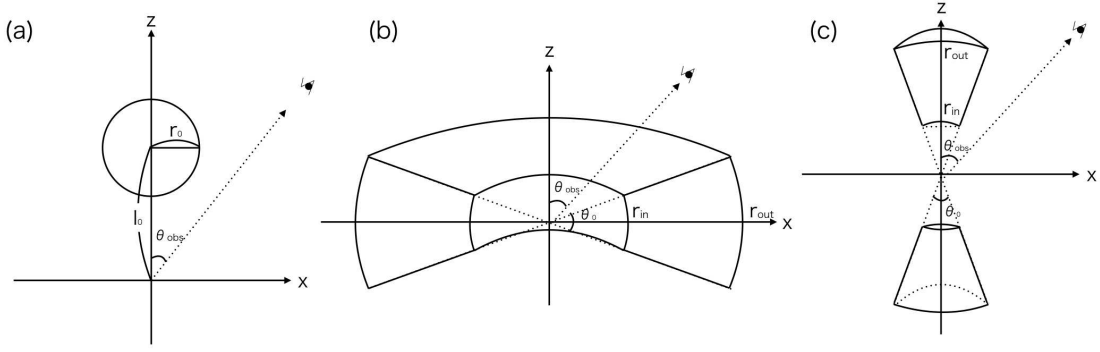}
\caption{Geometry of (a) the blob, (b) disk and (c) bipolar CS dust models.}
\end{center}
\end{figure*}

\subsubsection{The blob model}
In the blob model, a distance of the blob from the SN and a radius of the blob are denoted by $l_0$ and $r_0$, respectively. An observer's direction is expressed as an angle $\theta_{\mathrm{obs}}$, which is an angle between the observer's direction and a direction of the blob. We use several values for $l_0$: $1.5, 2.0, 2.5, 3.0, 3.5$ and $4.0 \times 10^{17}$ cm ($\sim 0.049, \sim 0.065, \sim 0.081, \sim 0.097, \sim 0.11$ and $\sim 0.13$ pc). The corresponding typical light travel time of the scattered light for each case is $l_0/c \times (1-\cos \theta_{\mathrm{obs}}) \sim 58, 77, 97, 116, 135$ and $154 \times (1-\cos \theta_{\mathrm{obs}})$ days, respectively (see Table 1). In cases where the optical depth of the blob is small, the delay time of the scattered light is almost equal to the light travel time. Otherwise, multiple scatterings within the blob increase the delay time, while this effect is generally weak as compared to the light travel time for the situations considered in this paper; to produce a detectable level of polarization, the optical depth should not exceed the unity substantially. Therefore, the light travel time provides a good measure of the delay time. In this study we roughly use the light travel time as the delay time.

\begin{table*}
\caption{The typical light travel time ([day]) in the blob models.}
\begin{tabular}{ccccccc}
\tableline
 $l_0 \times 10^{17}$ [cm] & light travel time [day] & $\theta_{\mathrm{obs}}=10$ deg & $\theta_{\mathrm{obs}}=30$ deg & $\theta_{\mathrm{obs}}=50$ deg & $\theta_{\mathrm{obs}}=70$ deg & $\theta_{\mathrm{obs}}=90$ deg\\

\tableline \tableline
1.5 & $\sim 58 (1-\cos \theta_{\mathrm{obs}})$  & $\sim 1$ & $\sim 8$  & $\sim 21$ & $\sim 38$  & $\sim 58$\\ \hline
2.0 & $\sim 77 (1-\cos \theta_{\mathrm{obs}})$  & $\sim 1$ & $\sim 10$ & $\sim 28$ & $\sim 51$  & $\sim 77$\\ \hline
2.5 & $\sim 97 (1-\cos \theta_{\mathrm{obs}})$  & $\sim 1$ & $\sim 13$ & $\sim 34$ & $\sim 64$  & $\sim 97$\\ \hline
3.0 & $\sim 116 (1-\cos \theta_{\mathrm{obs}})$ & $\sim 2$ & $\sim 16$ & $\sim 41$ & $\sim 76$  & $\sim 116$\\ \hline
3.5 & $\sim 135 (1-\cos \theta_{\mathrm{obs}})$ & $\sim 2$ & $\sim 18$ & $\sim 48$ & $\sim 89$  & $\sim 135$\\ \hline
4.0 & $\sim 154 (1-\cos \theta_{\mathrm{obs}})$ & $\sim 2$ & $\sim 21$ & $\sim 55$ & $\sim 102$ & $\sim 154$\\ \hline
\tableline
\end{tabular}
\end{table*}

The minimum value of $l_0$ is in principle determined by evaporation of CS dust by the initial ultraviolet (UV) flash after shock breakout, which is discussed in \S 4.2. The value of $r_0$ is set so that a covering fraction of the blob for the SN light is $0.01$ (i.e., the corresponding solid angle is $4 \pi \times 0.01$), which is further discussed in \S 4.3. Since the solid angle of the blob covering the SN light is $\pi r_{0}^{2} / l_{0}^{2}$, we set $r_0 = 0.2 l_0$. Density of the blob is assumed to be uniform within the blob ($\rho_0$). We adopt the following characteristic optical depth along a diameter of the blob ($\tau_0$) as our model parameter without specifying absolute values of the mass absorption coefficient ($\kappa_{\mathrm{ext},\nu}$) and density ($\rho_0$); $\tau_0 = 2 r_0 \kappa_{\mathrm{ext},\nu} \rho_0 = 0.1, 1.0, 2.0, 3.0$ and $10.0$. In the blob model, we therefore leave $l_0$ and $\tau_0$ as our tunable parameters.

In the blob model, total dust mass $M_{\mathrm{dust}}$ and the corresponding mass-loss rate $\dot{M}_{\mathrm{gas}}$, under an assumption that there is no mass loss except for the CS dusty blob, are derived as follows:
\begin{eqnarray}
M_{\mathrm{dust}} &=& \frac{4}{3}\pi r_{0}^{3} \rho_{0} = \frac{2\pi l_{0}^{2} \tau_0}{75 \kappa_{\mathrm{ext}}} \nonumber\\ 
&\sim& 7.5 \times 10^{-4} \Biggl( \frac{l_0}{3\times 10^{17}\; \mathrm{cm}} \Biggr)^{2} \Biggl( \frac{\tau_0}{2.0} \Biggr) \nonumber\\
&& \Biggl( \frac{\kappa_{\mathrm{ext}}}{10^{4}\; \mathrm{cm}^{2}\; \mathrm{g}^{-1}} \Biggr)^{-1}  \mathrm{M}_{\odot},\\
\dot{M}_{\mathrm{gas}} &\sim& \frac{M_{\mathrm{gas}}}{2 r_0 / v_{\mathrm{w}}} = \frac{\pi v_{\mathrm{w}} l_0 \tau_0}{15 f_{\mathrm{dust}} \kappa_{\mathrm{ext}}} \nonumber\\
&\sim& 2.0 \times 10^{-5} \Biggl( \frac{l_0}{3\times 10^{17}\; \mathrm{cm}} \Biggr) \Biggl( \frac{\tau_0}{2.0} \Biggr) \nonumber\\
&& \Biggl( \frac{\kappa_{\mathrm{ext}}}{10^{4}\; \mathrm{cm}^{2}\; \mathrm{g}^{-1}} \Biggr)^{-1} \Biggl( \frac{f_{\mathrm{dust}}}{0.01} \Biggr)^{-1} \nonumber \\
&& \Biggl( \frac{v_{\rm{w}}}{10^{6}\; \mathrm{cm}\; \mathrm{s}^{-1}} \Biggr) \; \mathrm{M}_{\odot} \mathrm{yr}^{-1},
\end{eqnarray}
where $v_{\mathrm{w}} $ and $f_{\mathrm{dust}}$ are wind velocity of a progenitor star and a dust-to-gas ratio, respectively. For $\kappa_{\mathrm{ext}}$, $f_{\rm{dust}}$ and $w_{\rm{w}}$, we adopt the values typically used in the literature \citep[e.g.,][]{Marshall2004, Draine2011, Mauron2011}, though the values are still observationally uncertain. The range of mass-loss rate for the range of $l_0$ and $\tau_0$ examined in this paper ($5.0 \times 10^{-7} \lesssim \dot{M} \lesssim 1.3 \times 10^{-4} \; \mathrm{M}_{\odot} \mathrm{yr}^{-1}$) is consistent with those observationally derived for RSGs \citep[$1.0 \times 10^{-7} \lesssim \dot{M} \lesssim 1.0 \times 10^{-4} \; \mathrm{M}_{\odot} \mathrm{yr}^{-1}$, e.g.,][]{Mauron2011}.

\subsubsection{The disk model}
In the disk model, the inner and outer radii of the disk are denoted by $r_{\mathrm{in}}$ and $r_{\mathrm{out}}$, respectively. An observer's direction is expressed as an angle $\theta_{\mathrm{obs}}$, which is an angle between the observer's direction and a polar direction of the disk. We use several values for $r_{\mathrm{in}}$: $1.5, 2.0, 2.5, 3.0, 3.5$ and $4.0 \times 10^{17}$ cm, while we set $r_{\mathrm{out}} = r_{\mathrm{in}} + 3.0 \times 10^{17}$ cm. An opening angle of the disk is denoted by $\theta_0$ and is set so that a covering fraction of the disk for the SN light is $0.01$. Since the solid angle of the disk covering the SN light is $4 \pi \sin (\theta_0 /2)$, we obtain $\theta_0 = 2 \arcsin (0.01) \sim 1.15$ degree. The radial density distribution of the CS dust is assumed to follow $\rho_{\mathrm{dust}} (r) = \rho_{\mathrm{dust}}(r_{\mathrm{in}}) (r/r_{\mathrm{in}})^{-2}$, as expected from a stationary mass loss from a progenitor star. We use the following optical depth along a disk plane as our model parameter: $\tau_0 = \kappa_{\mathrm{ext},\nu} \int_{r_{\mathrm{in}}}^{r_{\mathrm{out}}} \rho_{\mathrm{dust}} (r) dr = \kappa_{\mathrm{ext},\nu} \rho_{\mathrm{dust}} (r_{\mathrm{in}}) r_{\mathrm{in}}^2 (1/r_{\mathrm{in}} - 1/r_{\mathrm{out}}) = 0.1, 1.0, 2.0, 3.0$ and $10.0$. In the disk model, $r_{\mathrm{in}}$ and $\tau_0$ are our input parameters.

In the disk model, the total dust mass, $M_{\mathrm{dust}}$, and the corresponding mass-loss rate, $\dot{M}_{\mathrm{gas}}$, are derived as follows:
\begin{eqnarray*}
  M_{\mathrm{dust}} &=& \int_{r_{\mathrm{in}}}^{r_{\mathrm{out}}} 4 \pi r^2 \rho_{\mathrm{dust}} (r) dr \times \sin (\theta_0 /2) \nonumber \\
  &=& \frac{4\pi \sin (\theta_0 /2) r_{\mathrm{in}} r_{\mathrm{out}} \tau_0}{\kappa_{\mathrm{ext}}}\\
  &\sim& 8.5 \times 10^{-4} \Biggl( \frac{r_{\mathrm{in}}}{1.5\times 10^{17}\; \mathrm{cm}} \Biggr) \Biggl( \frac{r_{\mathrm{out}}}{4.5\times 10^{17}\; \mathrm{cm}} \Biggr) \nonumber \\
  && \Biggl( \frac{\tau_0}{2.0} \Biggr) \Biggl( \frac{\kappa_{\mathrm{ext}}}{10^{4}\; \mathrm{cm}^{2}\; \mathrm{g}^{-1}} \Biggr)^{-1} \; \mathrm{M}_{\odot},\\
  \dot{M}_{\mathrm{gas}} &=& 4 \pi r_{\mathrm{in}}^2 \rho_{\mathrm{gas}} (r_{\mathrm{in}}) v_{\mathrm{w}} \times \sin (\theta_0 /2) \nonumber \\
  &=& \frac{4 \pi \sin (\theta_0 /2) v_{\mathrm{w}} \tau_0}{f_{\mathrm{dust}} \kappa_{\mathrm{ext}}} \Biggl( \frac{r_{\mathrm{in}} r_{\mathrm{out}}}{r_{\mathrm{out}} - r_{\mathrm{in}}} \Biggr)\\
  &\sim& 8.9 \times 10^{-6} \Biggl( \frac{r_{\mathrm{in}}}{1.5\times 10^{17}\; \mathrm{cm}} \Biggr) \Biggl( \frac{r_{\mathrm{out}}}{4.5\times 10^{17}\; \mathrm{cm}} \Biggr) \nonumber \\
  && \Biggl( \frac{\tau_0}{2.0} \Biggr) \Biggl( \frac{\kappa_{\mathrm{ext}}}{10^{4}\; \mathrm{cm}^{2}\; \mathrm{g}^{-1}} \Biggr)^{-1} \Biggl( \frac{f_{\mathrm{dust}}}{0.01} \Biggr)^{-1} \nonumber \\
&& \Biggl( \frac{v_w}{10^{6}\; \mathrm{cm}\; \mathrm{s}^{-1}} \Biggr) \; \mathrm{M}_{\odot} \; \mathrm{yr}^{-1},
\end{eqnarray*}
where $v_{\mathrm{w}} $ and $f_{\mathrm{dust}}$ are wind velocity of a progenitor star and a dust-to-gas ratio, respectively.

\subsubsection{The bipolar CS dust model}
In the bipolar CS dust model, the inner and outer radii are denoted by $r_{\mathrm{in}}$ and $r_{\mathrm{out}}$, respectively. An observer's direction is expressed as an angle $\theta_{\mathrm{obs}}$, which is an angle between the observer's direction and the polar direction. We adopt the following values: $1.5, 2.0, 2.5, 3.0, 3.5$ and $4.0 \times 10^{17}$ cm for $r_{\mathrm{in}}$ and $r_{\mathrm{out}} = r_{\mathrm{in}} + 3.0 \times 10^{17}$ cm. An opening angle of the bipolar CS dust is denoted by $\theta_0$ and is set so that a covering fraction of the bipolar CS dust for the SN light is $0.01$. Since the solid angle of the bipolar CS dust is $4 \pi [1 -\cos (\theta_0 /2)]$, we adopt $\theta_0 = 2 \arccos (0.99) \sim 16.2$ degree. The radial density distribution of the CS dust is the same as the case for the disk model: $\rho_{\mathrm{dust}} (r) = \rho_{\mathrm{dust}}(r_{\mathrm{in}}) (r/r_{\mathrm{in}})^{-2}$. The characteristic optical depth, $\tau_0$, is adopted as the optical depth along the polar axis. In the bipolar CS dust model, the input parameters are $r_{\mathrm{in}}$ and $\tau_0$.

In the bipolar CS dust model, the total dust mass, $M_{\mathrm{dust}}$, and the corresponding mass-loss rate, $\dot{M}_{\mathrm{gas}}$, are given as follows:
\begin{eqnarray*}
  M_{\mathrm{dust}} &=& \int_{r_{\mathrm{in}}}^{r_{\mathrm{out}}} 4 \pi r^2 \rho_{\mathrm{dust}} (r) dr \times [1-\cos (\theta_0 /2)] \nonumber \\
  &=& \frac{4\pi [1-\cos (\theta_0 /2)] r_{\mathrm{in}} r_{\mathrm{out}} \tau_0}{\kappa_{\mathrm{ext}}}\\
  &\sim& 8.5 \times 10^{-4} \Biggl( \frac{r_{\mathrm{in}}}{1.5\times 10^{17}\; \mathrm{cm}} \Biggr) \Biggl( \frac{r_{\mathrm{out}}}{4.5\times 10^{17}\; \mathrm{cm}} \Biggr) \nonumber \\
  &&\Biggl( \frac{\tau_0}{2.0} \Biggr) \Biggl( \frac{\kappa_{\mathrm{ext}}}{10^{4}\; \mathrm{cm}^{2}\; \mathrm{g}^{-1}} \Biggr)^{-1} \; \mathrm{M}_{\odot},\\
  \dot{M}_{\mathrm{gas}} &=& 4 \pi r_{\mathrm{in}}^2 \rho_{\mathrm{gas}} (r_{\mathrm{in}}) v_{\mathrm{w}} \times [1-\cos (\theta_0 /2)] \nonumber \\
  &=& \frac{4 \pi [1-\cos (\theta_0 /2)] v_{\mathrm{w}} \tau_0}{f_{\mathrm{dust}} \kappa_{\mathrm{ext}}} \Biggl( \frac{r_{\mathrm{in}} r_{\mathrm{out}}}{r_{\mathrm{out}} - r_{\mathrm{in}}} \Biggr)\\
  &\sim& 8.9 \times 10^{-6} \Biggl( \frac{r_{\mathrm{in}}}{1.5\times 10^{17}\; \mathrm{cm}} \Biggr) \Biggl( \frac{r_{\mathrm{out}}}{4.5\times 10^{17}\; \mathrm{cm}} \Biggr) \nonumber \\
  &&\Biggl( \frac{\tau_0}{2.0} \Biggr) \Biggl( \frac{\kappa_{\mathrm{ext}}}{10^{4}\; \mathrm{cm}^{2}\; \mathrm{g}^{-1}} \Biggr)^{-1} \Biggl( \frac{f_{\mathrm{dust}}}{0.01} \Biggr)^{-1} \nonumber \\
&& \Biggl( \frac{v_w}{10^{6}\; \mathrm{cm}\; \mathrm{s}^{-1}} \Biggr) \; \mathrm{M}_{\odot} \; \mathrm{yr}^{-1}.
\end{eqnarray*}
The values of $M_{\mathrm{dust}}$ and $\dot{M}_{\mathrm{gas}}$ are the same with those in the disk model.

\subsection{Input SN light}
As an input SN light curve, we use a simple light curve that mimics the observed optical light curve of SNe IIP. In this paper, we do not specify an exact wavelength for the input SN light but suppose it to be in the optical bands, which is further discussed in \S 4.4. The absolute magnitude of the SN is assumed to be $-16$ mag until 85 days since the explosion (the plateau phase), $-13.5$ mag after 120 days (the nebular phase) and the linearly-interpolated values between the two phases (85-120 days). The assumption of the constant luminosity in each phase (the plateau and nebular phases) is further discussed in \S 4.3. In the optical wavelengths, these values are typical for SNe IIP \citep[e.g.,][]{Anderson2014, Sanders2015}. The SN light is assumed to be unpolarized. A source of polarization is only dust scattering. In this study, photons are emitted from the origin without taking into account the expansion of the SN photosphere. The size of the SN photosphere is generally negligible; it is derived as $(L_{\rm{SN}}/4 \pi \sigma_{\rm{SB}} T_{\rm{I}}^{4})^{1/2} < 1.0$ light days, where $L_{\rm{SN}}$ is the typical SN luminosity ($\sim 10^{42}$ erg s$^{-1}$), $\sigma_{\rm{SB}}$ is the Stefan-Boltzmann constant and $T_{\rm{I}}$ is the ionization temperature of hydrogen ($\lesssim 6000$ K).

\section{results}
\subsection{general properties}
First we discuss the general behavior of polarization in the dust scattering model, providing predictions for each of the three models. 

\subsubsection{The blob model}
Figure 2 shows time evolution of the polarization degree $P$ in the blob model for various values of $\theta_{\mathrm{obs}}$ ($\theta_{\mathrm{obs}} = 10, 30, 50, 70$ and $90$ degree). Two cases are shown for $l_{0} = 2.0$ and $4.0 \times 10^{17}$ cm. The corresponding light travel time for each $\theta_{\mathrm{obs}}$ is shown in Table 1. For the case with $\theta_{\mathrm{obs}} = 10$ degree, the polarization degree is zero regardless of a value of $l_0$. This is because the linear polarization by the scattered echoes from CS dust is almost totally cancelled out due to the projected circular symmetry as viewed from an observer (see Figure 3). Figure 2a shows the results for the case with $l_0 = 2.0 \times 10^{17}$ cm. In this case, the delay time is shorter than $85$ days irrespective of $\theta_{\mathrm{obs}}$. Initially, the polarization degree slightly rises when the polarized-scattered echo reaches an observer in the timescale of the delay time (in the plateau phase). Then, the polarization degree is rapidly increased as the SN becomes fainter and, thus, the contribution of the echo becomes higher (in the decline phase). After reaching to the maximum ($P=P_{\mathrm{max}}$), the polarization degree becomes decreased back to the same level with the first rise as the contribution of the echo also becomes lower (in the nebular phase). The value of $P_{\mathrm{max}}$ is determined by the relative flux of the polarized-scattered echo from the blob to the unpolarized SN flux. If the delay time of the scattered echo is longer than the timescale of the SN flux decrease (i.e., $(120-85)=35$ days in the current situation), which is the case for $\theta_{\mathrm{obs}} = 70$ and $90$ degree, the relative flux can (roughly) reach to the maximum value as described above. On the other hand, if the delay time of the scattered echo is shorter than $35$ days, which is the case for $\theta_{\mathrm{obs}} = 30$ and $50$ degree, the flux of the scattered echo is already somewhat decreased before the SN becomes faint enough to reach to the tail phase ($-13.5$ mag). Therefore, the values of $P_{\mathrm{max}}$ are lower than those in the other cases. 

\begin{figure}[t]
  \includegraphics[scale=0.7]{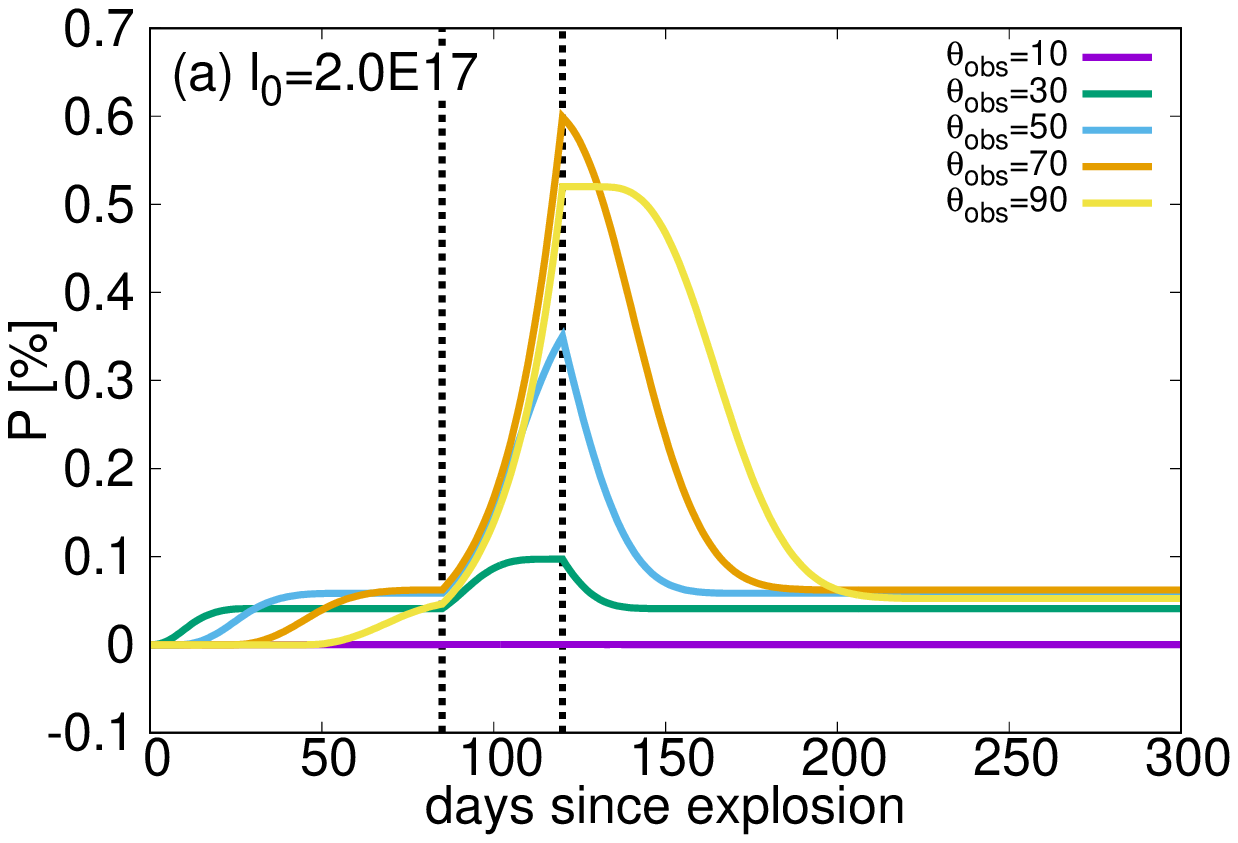}
  \includegraphics[scale=0.7]{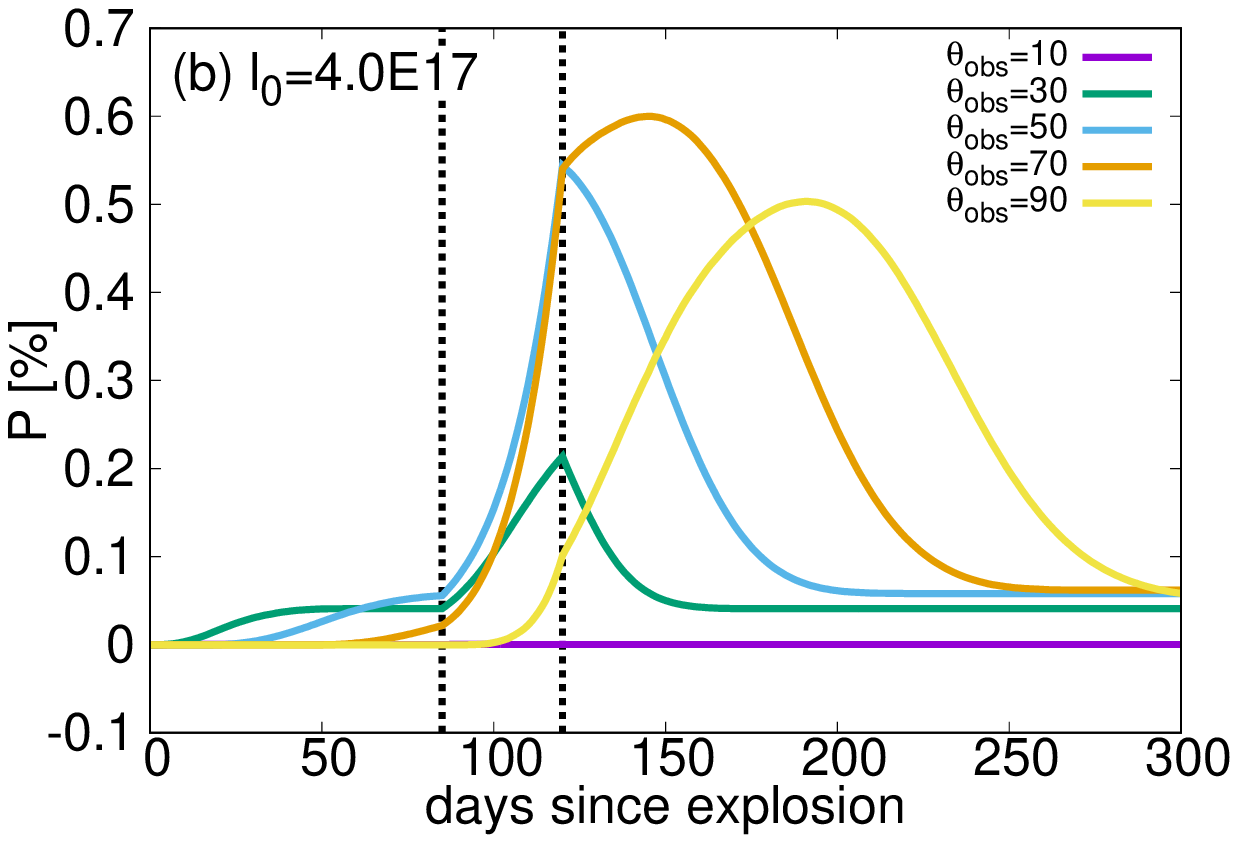}
\caption{(a) Time evolution of the polarization degree in the blob model for various values of $\theta_{\mathrm{obs}}$ [degree], where $\tau_0 = 2.0$ and $l_0 = 2.0 \times 10^{17}$ cm. (b) Same as (a), but for $l_0 = 4.0 \times 10^{17}$ cm. The vertical dotted lines show the epochs that separate the three different phases (i.e. $85$ and $120$ days, see \S 2.3).}
\end{figure}

\begin{figure}[t]
  \includegraphics[scale=0.23]{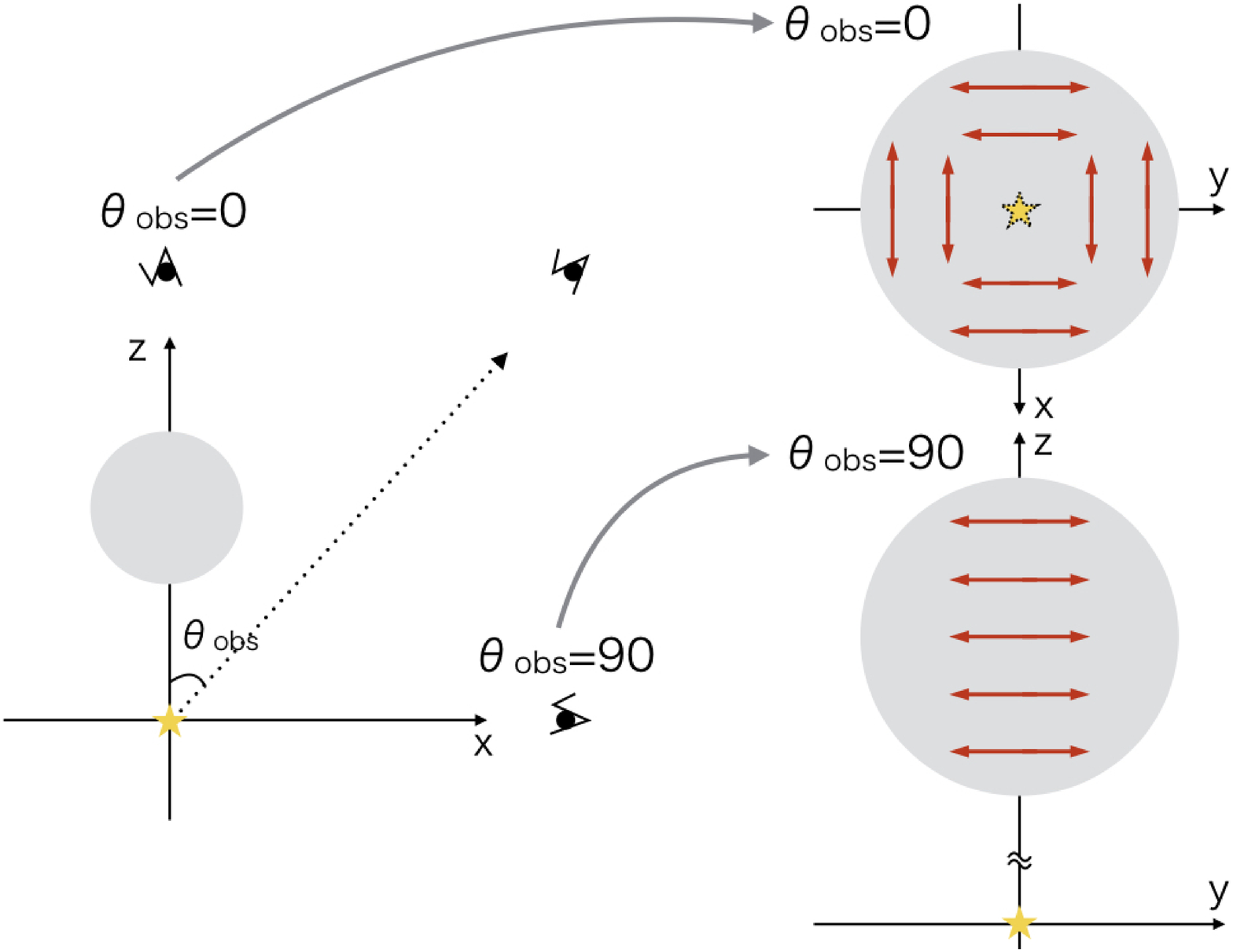}
\caption{Schematic picture of the observed polarization for the cases with $\theta_{{\rm obs}}=0$ and 90 degree in the blob model.}
\end{figure}

The flux and polarization degree of the scattered echo depend on $\theta_{\mathrm{obs}}$. The flux of the echo is maximized at $\theta_{\mathrm{obs}} \sim 0$ degree for the same input flux, because we adopt $g=0.6$ as a scattering asymmetry parameter, which means that a scattering process is a relatively forward scattering. The polarization degree of the individual echo components are maximized at $\theta_{\mathrm{obs}} \sim 90$ degree (see Figure 3). Thus, the contribution of the echo for SN polarization is maximized for an intermediate angle, and particularly around $\theta_{\mathrm{obs}} \sim 50$ degree in our simulations. However, the difference of the contributions of the echo for SN polarization is less than $\sim 30$ \% for $30 \lesssim \theta_{\mathrm{obs}} \lesssim 90$ degree, under our dust model parameters \citep[see][]{Code1995}. Therefore, the value of $P_{\mathrm{max}}$ does not sensitively depends on $\theta_{\mathrm{obs}}$ for $30 \lesssim \theta_{\mathrm{obs}} \lesssim 90$ degree, as long as the delay time of the scattered echo is longer than 35 days, as is the case with $\theta_{\mathrm{obs}} \gtrsim 70$ in Figure 2a.

Figure 2b shows the results for the case with $l_0 = 4.0 \times 10^{17}$ cm. In this case, the time evolution is similar to the case that has the similar delay time in the cases with $l_0 = 2.0 \times 10^{17}$ cm. In the case with $\theta_{\mathrm{obs}} = 90$ degree, where the delay time is longer than 85 days, the timing of polarization rise is delayed as compared to the luminosity drop of the SN. Therefore, the cases whose delay time is longer than 85 days are not preferred to explain the rise of the polarization synchronized to the SN luminosity drop as observed for some SNe IIP. It should also be noted that, for lager $l_0$, the effect of the light-crossing time within a blob becomes also important to determine the time evolution as is seen in the case with $l_0 = 4.0 \times 10^{17}$ cm.

The value of $P_{\mathrm{max}}$ is not highly dependent on $l_0$, if the corresponding delay time is larger than 35 days. Figure 4 shows time evolution of the polarization degree $P$ in the blob model for $\theta_{\mathrm{obs}} = 70$ degree and $\tau_0 = 2$, calculated for various values of $l_0$ ($l_0 = 1.5, 2.0, 2.5, 3.0, 3.5$ and $4.0 \times 10^{17}$ cm). Timing of $P_{\mathrm{max}}$ is slightly delayed as $l_0$ become larger, due to the effect of light-crossing time within a blob.

\begin{figure}[t]
  \includegraphics[scale=0.7]{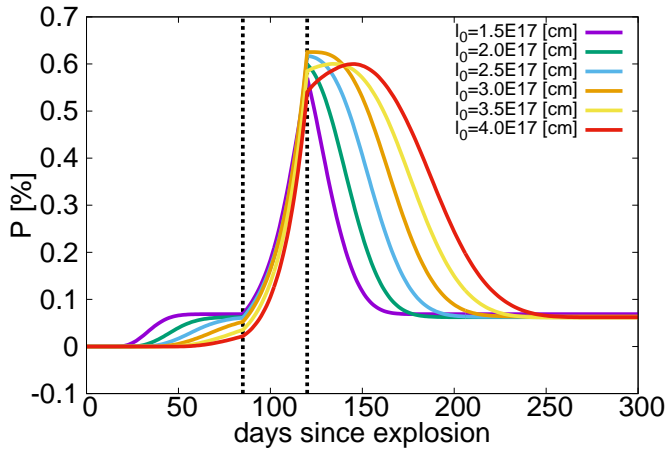}
\caption{Time evolution of the polarization degree in the blob model for various values of $l_0$, where $\tau_0 = 2.0$ and $\theta_{\mathrm{obs}} = 70$ degree. The vertical dotted lines are the same as Figure 2.}
\end{figure}

The shape of the time evolution of the polarization is not largely affected by $\tau_0$. Figure 5 shows time evolution of the polarization degree $P$ in the blob model for various values of $\tau_0$. The shape of the time evolution of the polarization is not dependent on $\tau_0$, except for the case with $\tau_0 = 10$. In the case with a high value of $\tau_0$, the polarization degree drops more rapidly than the other cases, because photons from the far side of a blob toward an observer, which are responsible for the polarization at later time, are selectively absorbed by dust in the blob. The value of $P_{\mathrm{max}}$ turns out to be largest for $\tau_0 = 2-3$. When $\tau_0 \lesssim 2-3$, the flux of the scattered echo, therefore the value of $P_{\mathrm{max}}$, are proportional to the value of $\tau_0$. When $\tau_0  \gtrsim 2-3$, the polarization degree of the scattered echo becomes lower as the value of $\tau_0$ becomes higher, due to increasing importance of multiple scattering.

\begin{figure}[t]
  \includegraphics[scale=0.7]{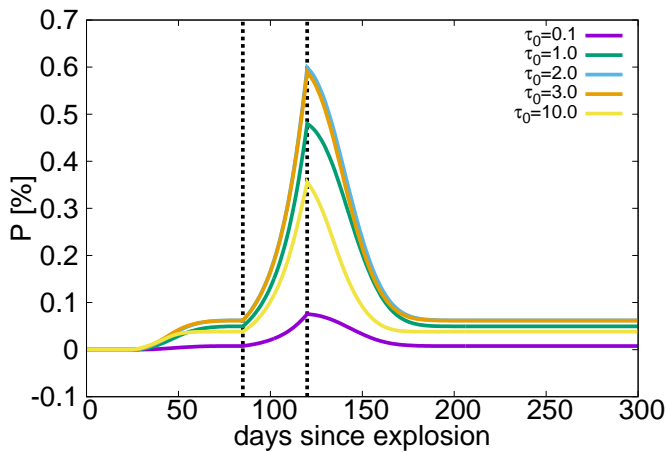}
\caption{Time evolution of the polarization degree in the blob model for various values of $\tau_0$, where $\theta_{\mathrm{obs}} = 70$ degree and $l_0 = 2.0 \times 10^{17}$ cm.}
\end{figure}

Figure 6 shows the values of $P_{\mathrm{max}}$ and $\Delta t$ for the blob models for various values of $l_0$ and $\tau_0$ toward $\theta_{\mathrm{obs}} = 70$ degree, where we define the duration of the high polarization ($\Delta t$) as the time duration from the point when the polarization becomes higher than $0.5 P_{\mathrm{max}}$ to the point when the polarization again becomes lower than $0.5 P_{\mathrm{max}}$ (full width at half maximum in the time-evolution curves of the polarization). The value of $P_{\mathrm{max}}$ basically depends on $\tau_0$. The value is peaked at $\tau_0 \sim 2.5$ ($P_{\mathrm{max}} \sim 0.65$ \%). The value of $\Delta t$ depends on $l_0$ and slightly on $\tau_0$.

\begin{figure}[t]
  \includegraphics[scale=0.7]{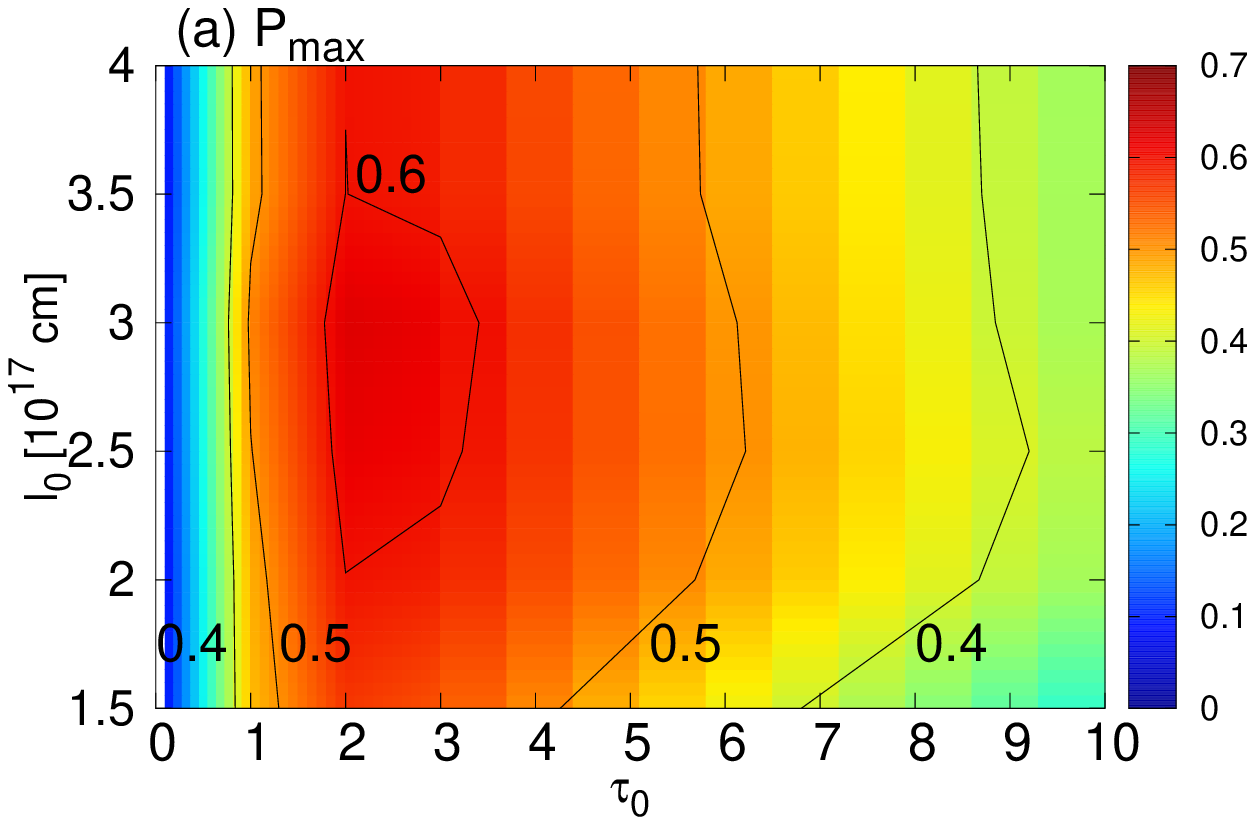}
  \includegraphics[scale=0.7]{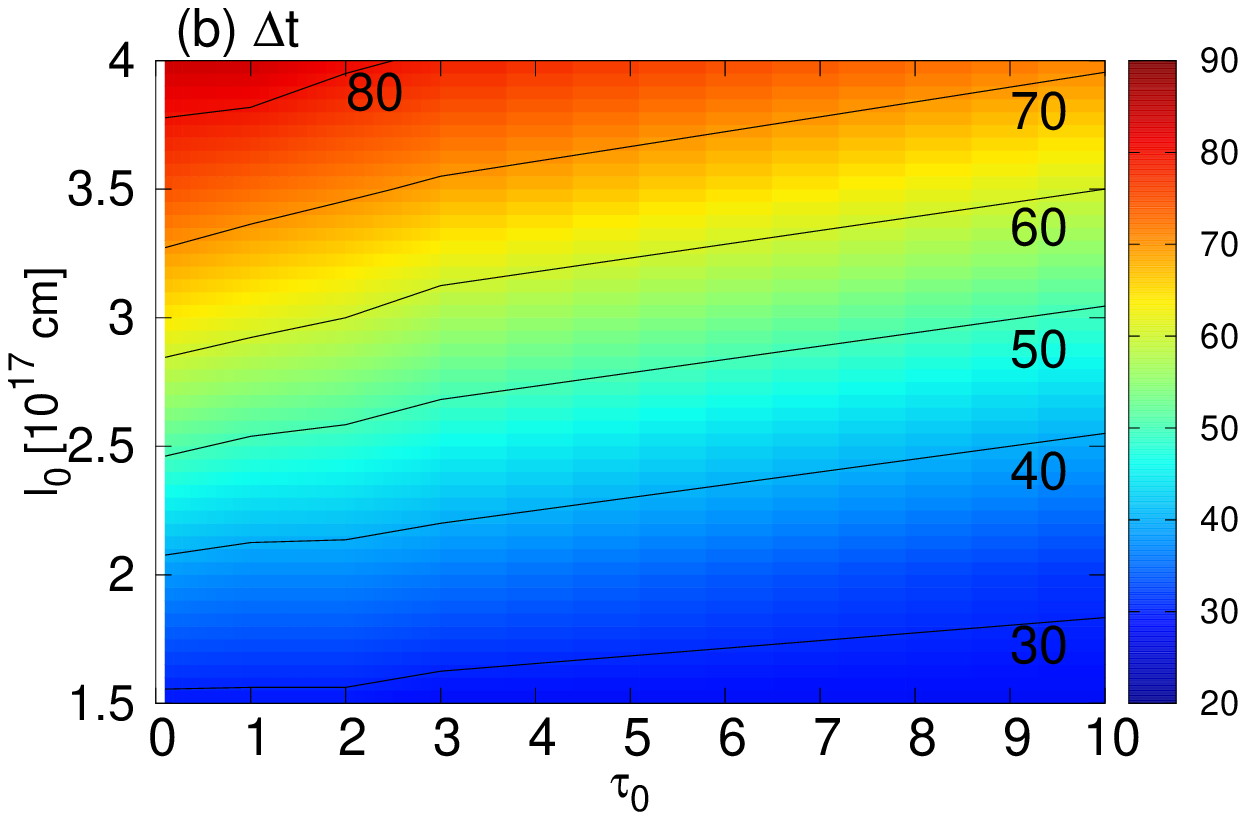}
\caption{The color and the black contour lines show values of (a) $P_{\mathrm{max}}$ [\%] and (b) $\Delta t$ [day] for the blob models for various values of $\tau_0$ and $l_0$, where $\theta_{\mathrm{obs}} = 70$ degree.}
\end{figure}

\subsubsection{The disk model}
In the disk model, the time evolution of polarization is different from that in the blob model. Figure 7 shows the same quantities as Fig. 5, but for the disk model for $\theta_{\mathrm{obs}} = 30$ degree (Fig. 7a) and $\theta_{\mathrm{obs}} = 70$ degree (Fig. 7b), for various values of $r_{\mathrm{in}}$. In the disk model, the position angle of the polarization depends on the part of the disk from which the scattered echo is originated. Assuming the single scattering case, photons from the closest (and farthest) side of the disk (the first (and third) component) have a position angle that is parallel to the $y$ axis for the observer with $(r,\theta,\phi)=(\infty,\theta_{\mathrm{obs}},0)$ in the spherical coordinates, while the other photons (the second componet) have the vertical component (see Fig. 8). Due to the different delay time of photons coming from different regions, the position angle of the liner polarization temporally evolves in the disk model. Generally, the third component does not contribute to the polarization, because its flux and polarization degree are small since they are created by the back scatterings. The relative flux and delay time of each component depend on $\theta_{\mathrm{obs}}$. The area where the first component is originated is smaller for larger $\theta_{\mathrm{obs}}$ (see Fig. 8), and thus the effective emitting point of the second component is closer to an observer. Therefore, the relative flux and delay time of the second component is larger and shorter, respectively, for larger $\theta_{\mathrm{obs}}$. The two polarization peaks in Fig. 7a represent these components. Initially, the first component is dominant due to the shorter delay time. Then, the second component gradually overwhelms the first component. The value of $P$ becomes zero when their contributions become comparable with each other. Finally, the second component becomes dominant due to the higher flux and polarization degree. In the disk model for $\theta_{\mathrm{obs}} = 70$ degree, the first component does not contribute to the time evolution of the polarization in SNe IIP, because the area where the first component is created is small and the delay time of the second component is short.

\begin{figure}[thbp]
  \includegraphics[scale=0.7]{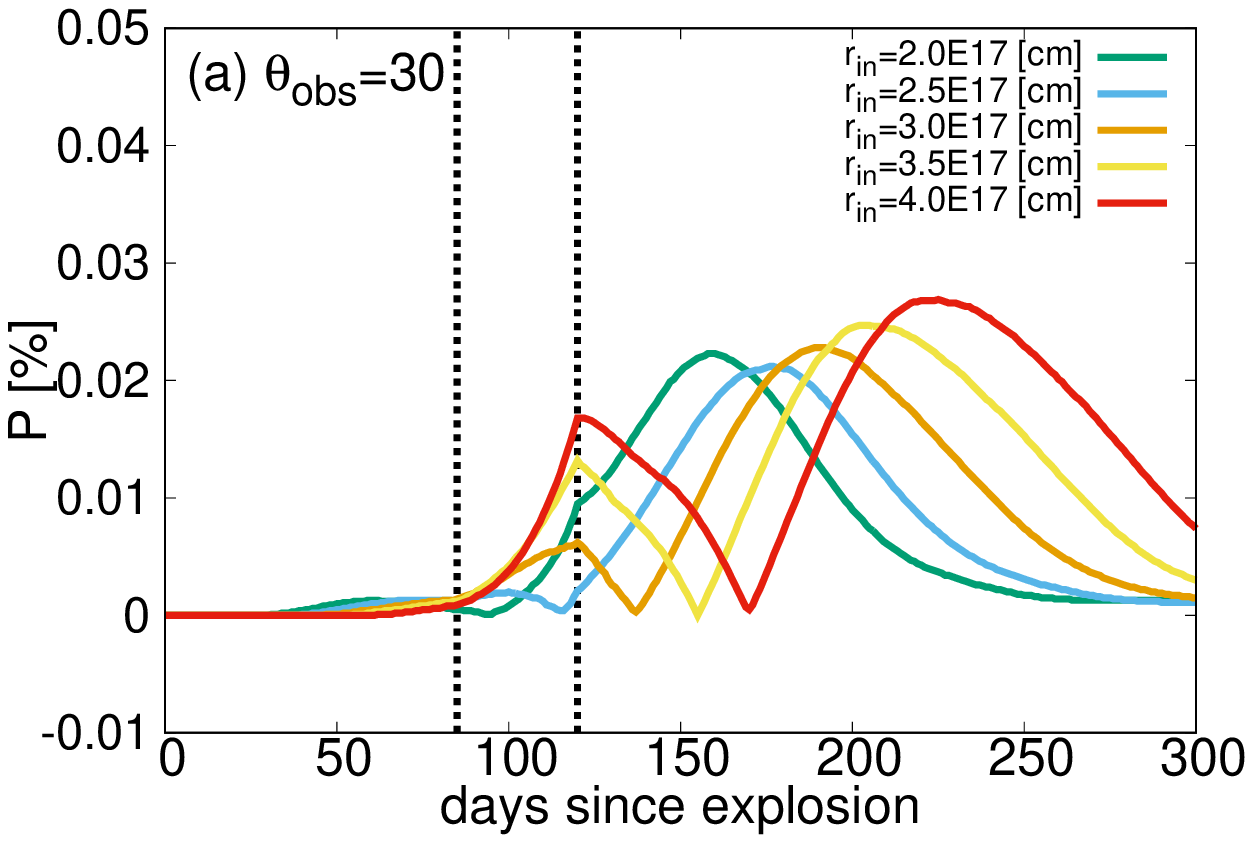}
  \includegraphics[scale=0.7]{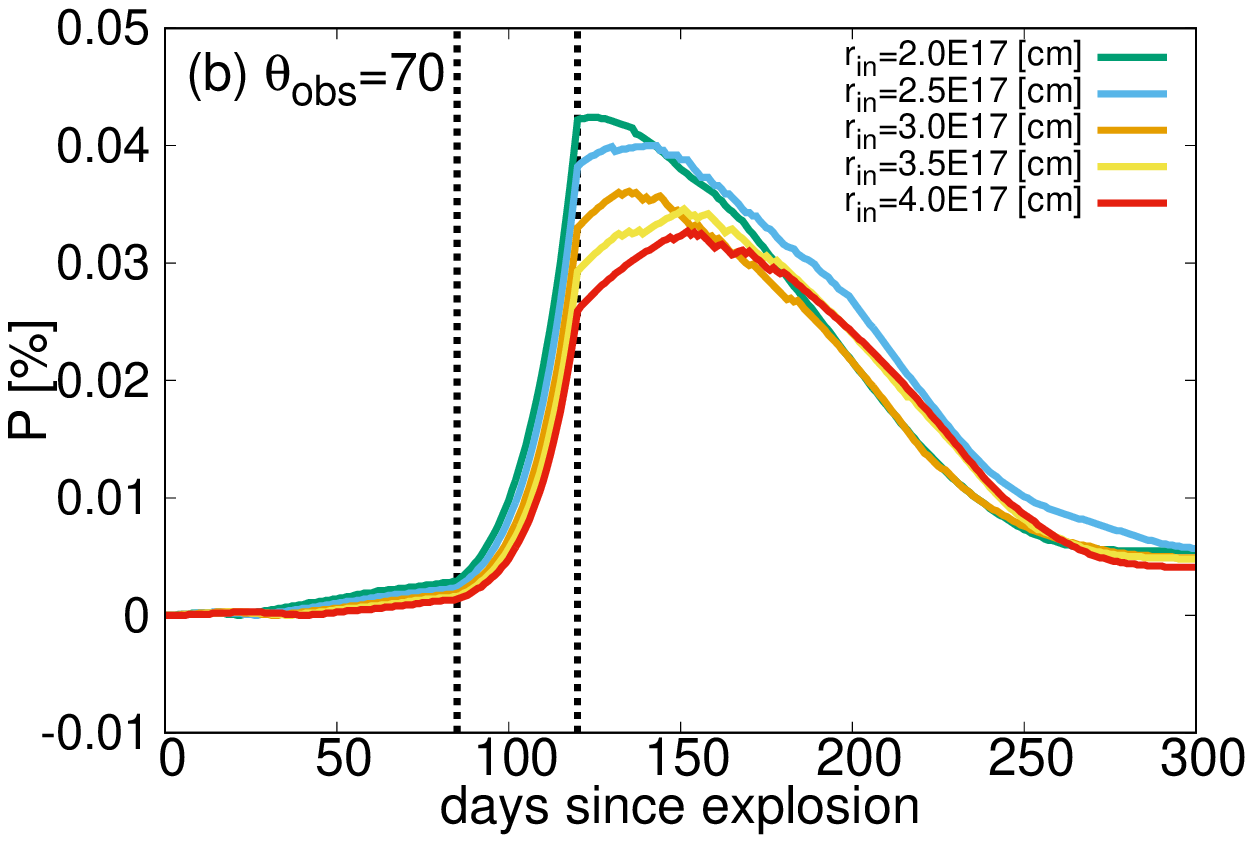}
\caption{Same as Fig. 4, but for the disk model for (a) $\theta_{\mathrm{obs}} = 30$ and (b) $\theta_{\mathrm{obs}} = 70$ degree, where $\tau_0 = 2.0$.}
\end{figure}

\begin{figure}[t]
  \centering
  \includegraphics[scale=0.25]{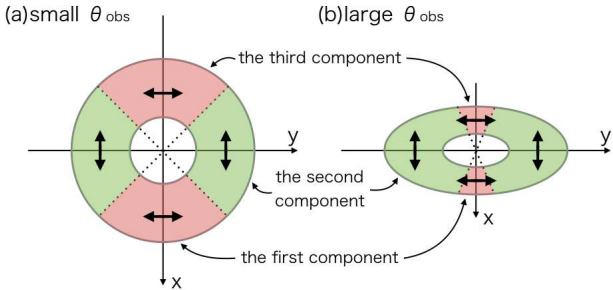}
\caption{Schematic picture of the position angle of the polarization in the disk model.}
\end{figure}

As shown in Figures 7a and 7b, the values of $P_{\mathrm{max}}$ are smaller than those in the blob model. This is mainly because of the axially symmetric structure of the disk. The closest side to the observer produces the polarization parallel to the y axis. When one moves to the farther region in the projected disk structure, the perpendicular component becomes stronger. Therefore, the polarizations from different regions tend to cancel out. If we make the opening angle of the disk larger for given $\tau_0$ (the amount of the CS dust higher), we may get higher polarization degree in the disk model (also discussed in \S4.3).

In the disk model, there is no case that can make the value of $P_{\mathrm{max}}$ larger than $0.1$ \%. Figure 9 shows the values of $P_{\mathrm{max}}$ in the disk models for various values of $l_0$ and $\tau_0$ toward $\theta_{\mathrm{obs}} = 70$ degree. We conclude that the disk model cannot explain the observed polarization feature in SNe IIP (typically $P_{\rm{max}} \sim 1$ \%, see \S 1) for a range of the parameters adopted in this study.

\begin{figure}[t]
  \includegraphics[scale=0.7]{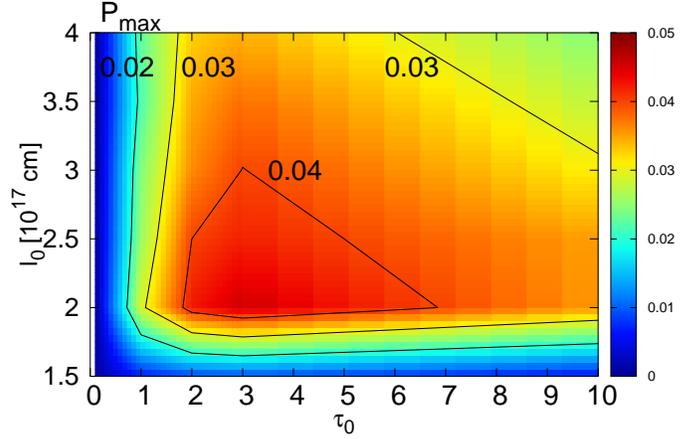}
\caption{Same as Fig. 6a, but for the disk model.}
\end{figure}

\subsubsection{The bipolar CS dust model}
The time evolution of the polarization in the bipolar CS dust model is similar to the case in the blob model. Figure 10 shows the same quantities as Fig. 5, but for the bipolar CS dust model. Since the difference of the light travel time for each part of bipolar CS dust is larger than that of the blob model, the polarization curves become wider in time. The general behavior for the bipolar CS dust model can be understand in the similar way as the blob model. Figure 11 shows the values of $P_{\mathrm{max}}$ and $\Delta t$ in the bipolar CS dust models for various values of $l_0$ and $\tau_0$ toward $\theta_{\mathrm{obs}} = 70$ degree.

\begin{figure}[t]
  \includegraphics[scale=0.7]{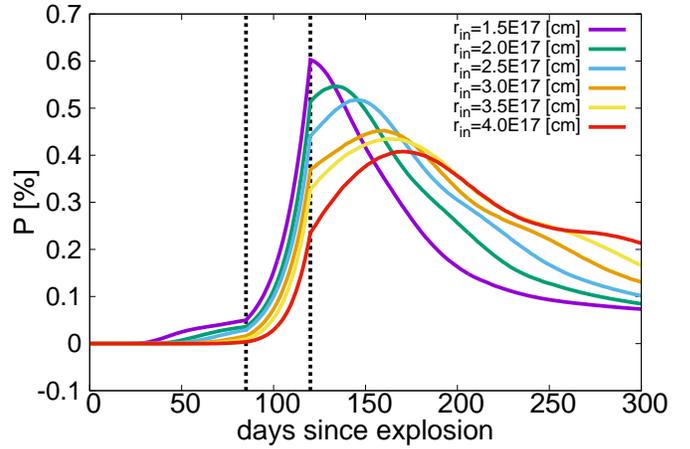}
\caption{Same as Fig. 4, but for the bipolar CS dust model, where $\tau_0 = 2.0$ and $\theta_{\mathrm{obs}} = 70$ degree.}
\end{figure}

\begin{figure}[t]
  \includegraphics[scale=0.7]{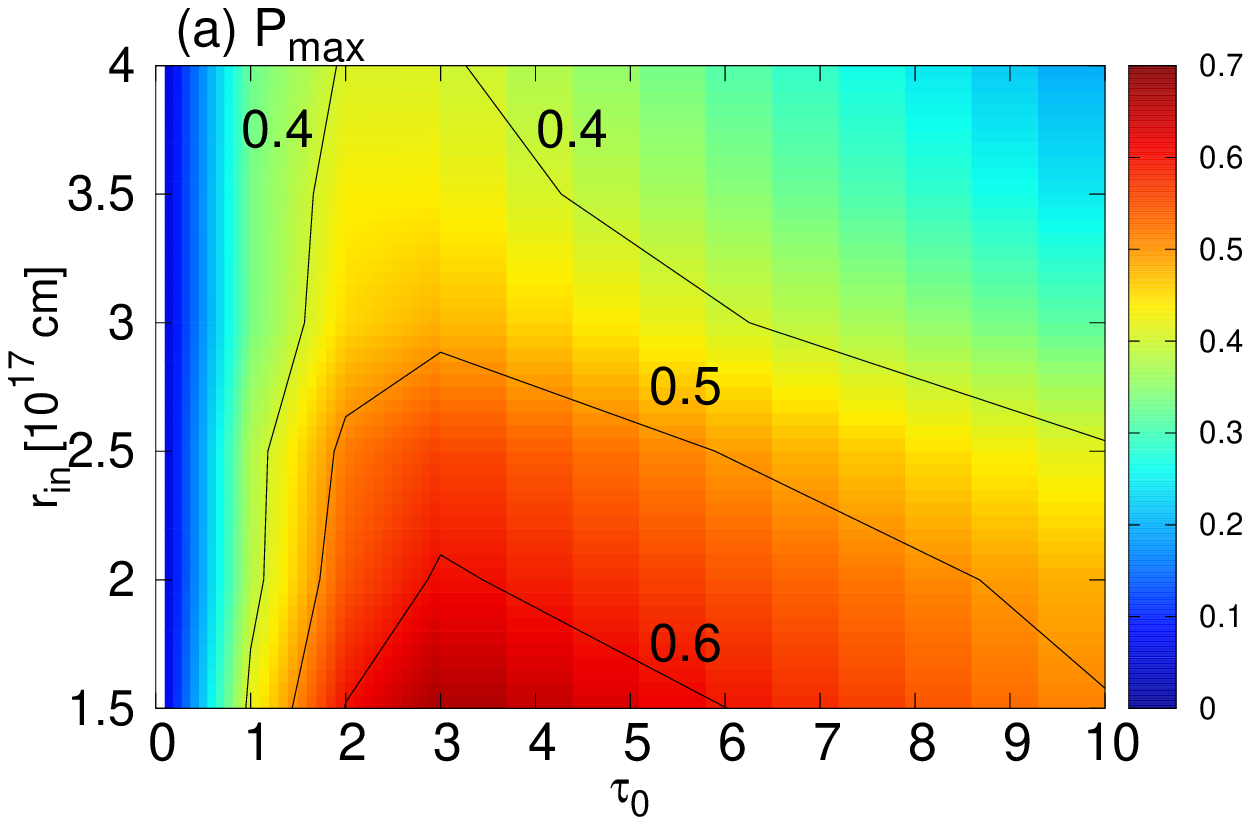}
  \includegraphics[scale=0.7]{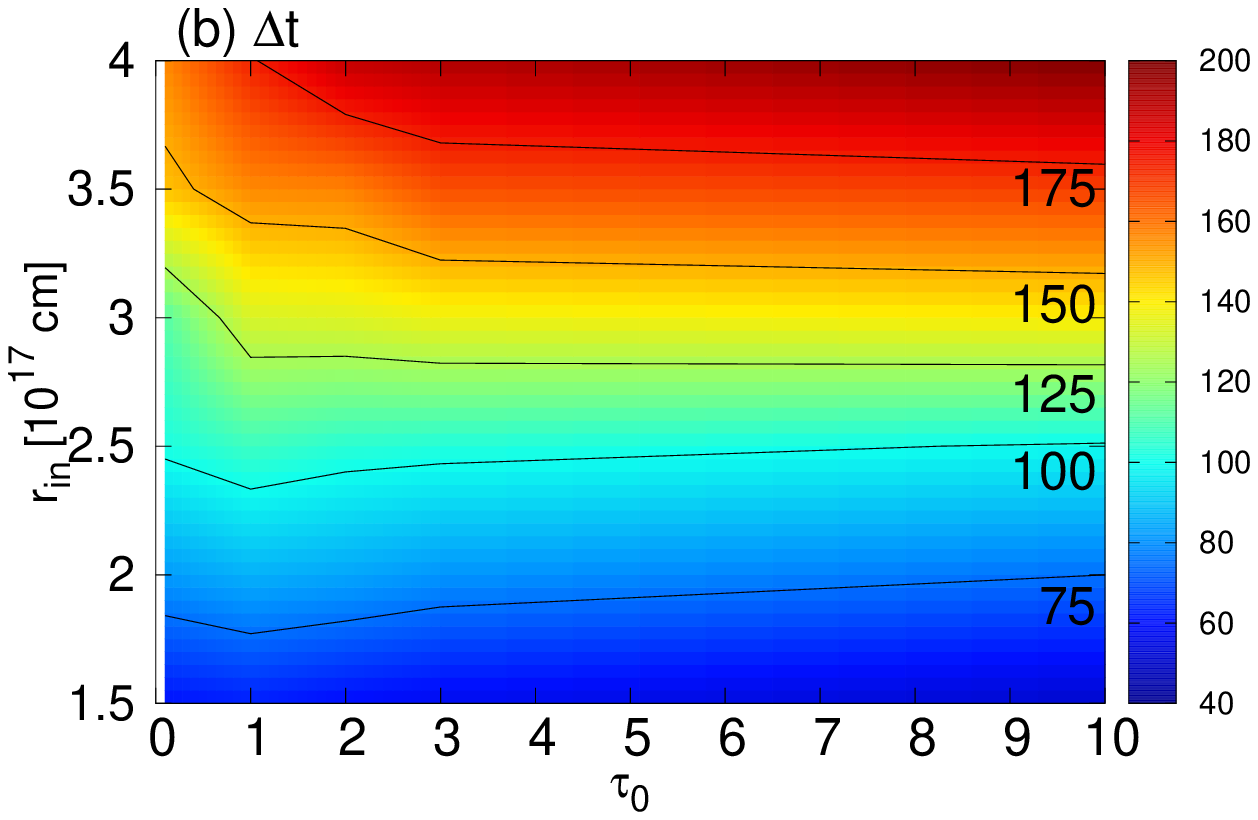}
\caption{Same as Fig. 6, but for the bipolar CS dust model.}
\end{figure}

In the blob and bipolar CS dust models, the position angle of the linear polarization $\chi$ is calculated to be steady and always perpendicular to a vector from the SN to the blob (or bipolar CS dust) on the sky (see Fig.3). The position angle in the disk models for some values of $\theta_{\mathrm{obs}}$ is also steady at first, but rotates by 90 degree at a certain time, as the dominant source points of the scattered echo in the disk are changing due to the delay time. This feature (not only for the disk configuration but also, i.e., for a two-blob configuration) may be important to reveal the geometry of CS dust, even though the polarization degree in the disk model is not large enough.

\subsection{Applications to SNe 2004dj and 2006ov}
In this section, we compare our results to observations of polarization in SNe IIP. Such observations are still rare. Here we focus on SNe 2004dj and 2006ov \citep[][]{Leonard2006, Chornock2010}, for which densely sampled polarimetric observations that allow to determine $P_{\mathrm{max}}$ and $\Delta t$ are available.

The best observed case for polarization of SNe IIP is for SN 2004dj, which shows the polarization characterized by $P_{\mathrm{max}} \sim 0.6$ \% and $\Delta t \sim 60$ days for wavelengths around $\sim 7000-8000 \AA$ \citep{Leonard2006}. The luminosity drop of SN 2004dj in the $R$ band is $\sim 2.5$ mag, which is the same value we assumed in this study. Figures 12a and 12b show the values of $P_{\mathrm{max}}$ and $\Delta t$ in SN 2004dj, compared with the values in the blob and bipolar CS dust models for various values of $\tau_0$ and $l_0$ with $\theta_{\mathrm{obs}} = 70$ degree. From the comparison between the observation and the models, it is found that the values of $P_{\mathrm{max}}$ and $\Delta t$ in SN 2004dj can be reproduced using the blob model for $l_0 \sim 3.0 \times 10^{17}$ cm, $\tau_0 \sim 2.0$ and $\theta_{\mathrm{obs}} = 70$ degree, or the bipolar CS dust model for $r_{\mathrm{in}} \sim 1.5 \times 10^{17}$ cm, $\tau_0 \sim 2.0$ and $\theta_{\mathrm{obs}} = 70$ degree. Figures 13a and 13b show the time evolution of the observed polarization in SN 2004dj, compared with those in the blob and bipolar CS dust models for the best-fit parameters derived above. They show a good match with our CS dust models.

\begin{figure}[t]
  \includegraphics[scale=0.7]{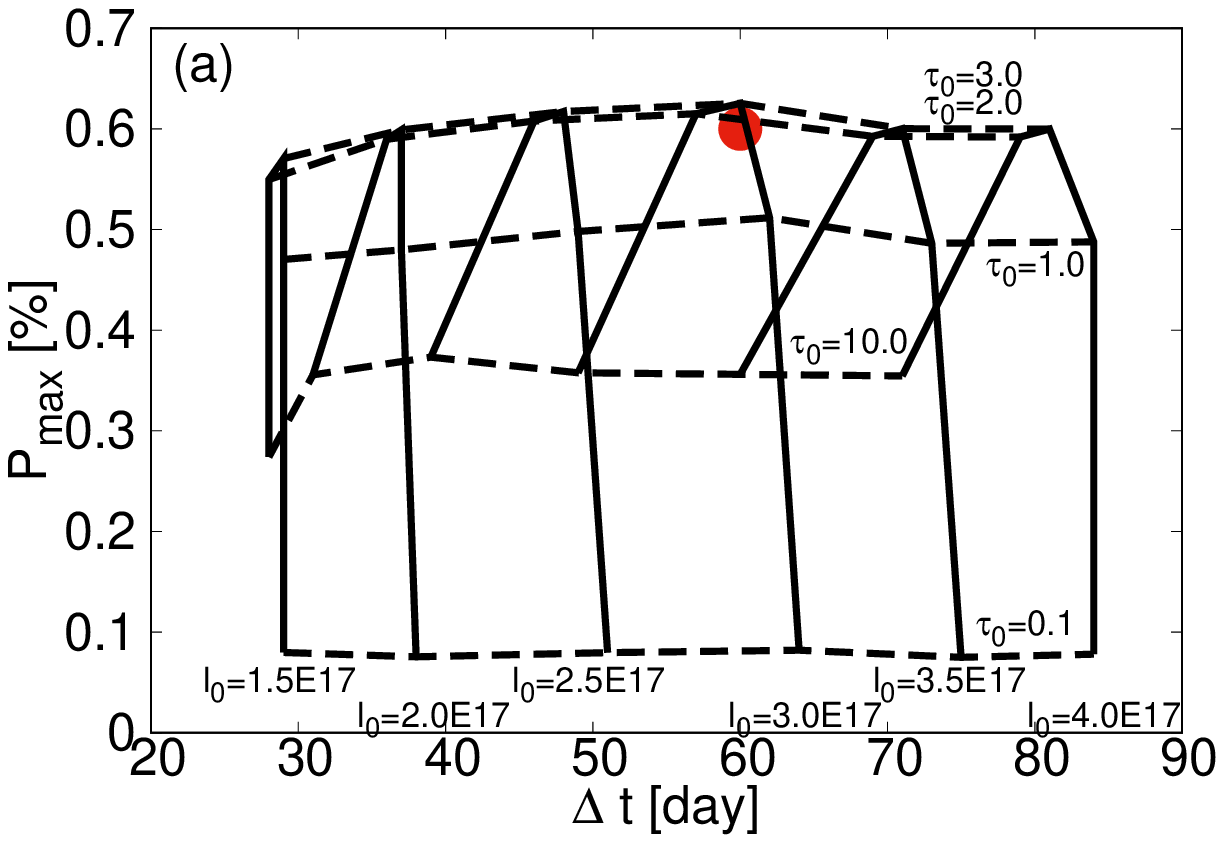}
  \includegraphics[scale=0.7]{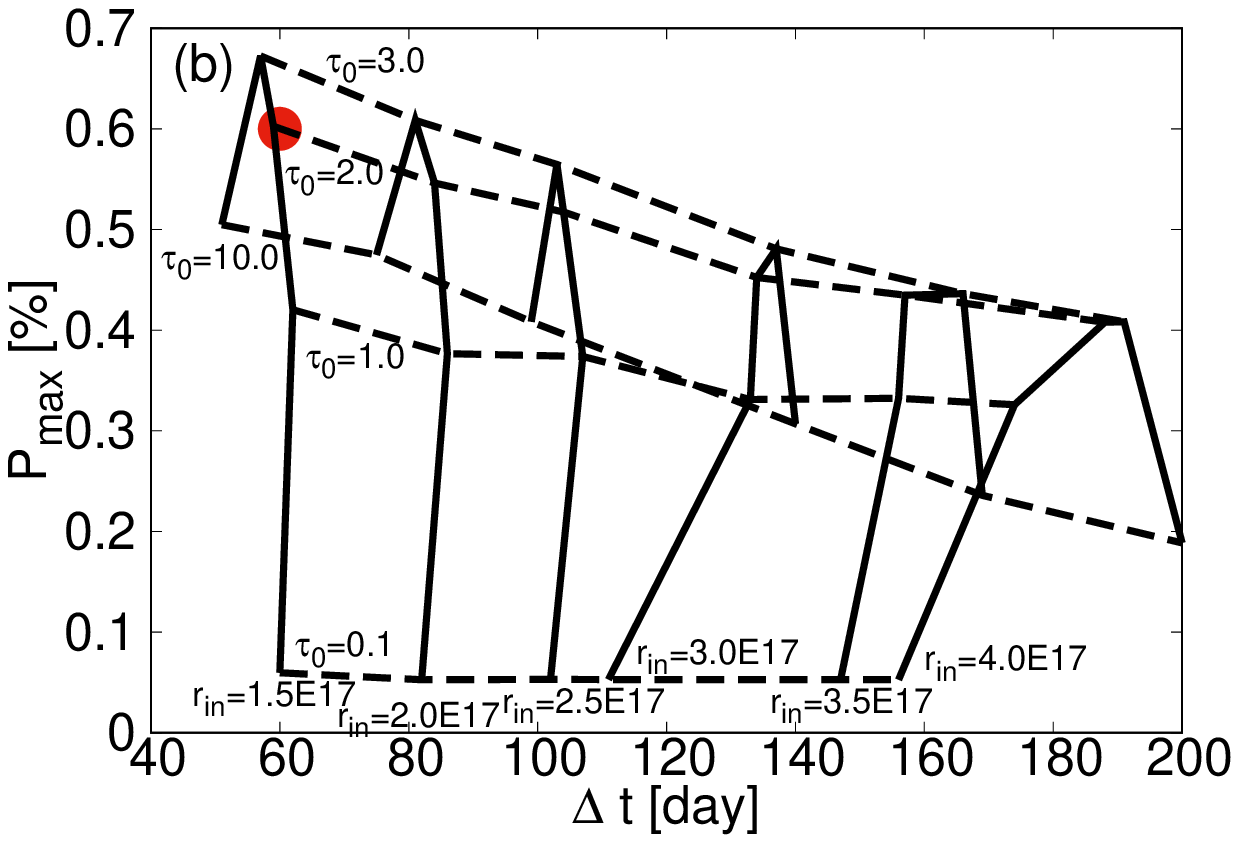}
\caption{(a) Values of $P_{\mathrm{max}}$ and $\Delta t$ in SN 2004dj (the red points; Leonard et al. 2006), compared with the values in the blob models with $\theta_{\rm{obs}}=70$ degree for various values of $\tau_0$ and $l_0$. The solid and dashed contour lines show the values in the blob models for various values of $l_0$ and $\tau_0$, respectively. (b) Same as (a), but for the bipolar CS dust models.}
\end{figure}

\begin{figure}[t]
  \includegraphics[scale=0.7]{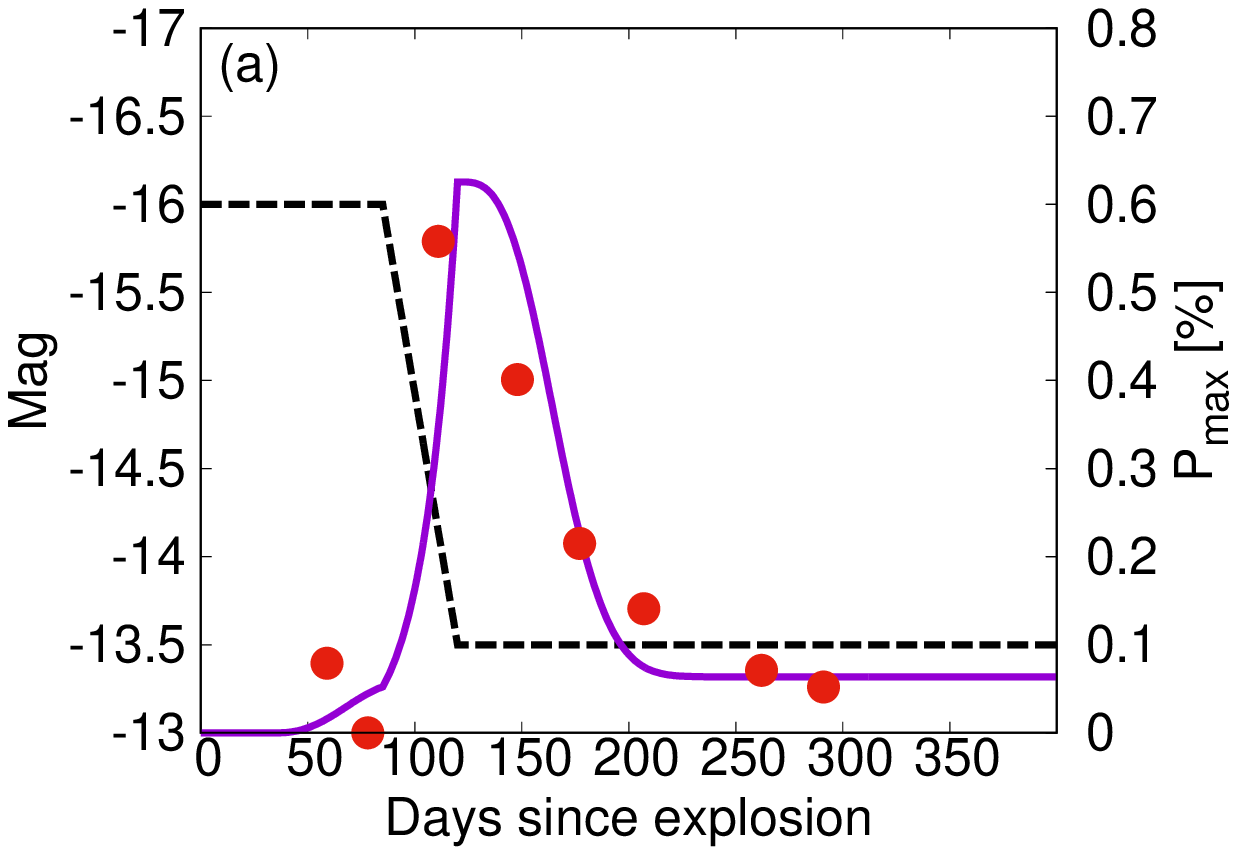}
  \includegraphics[scale=0.7]{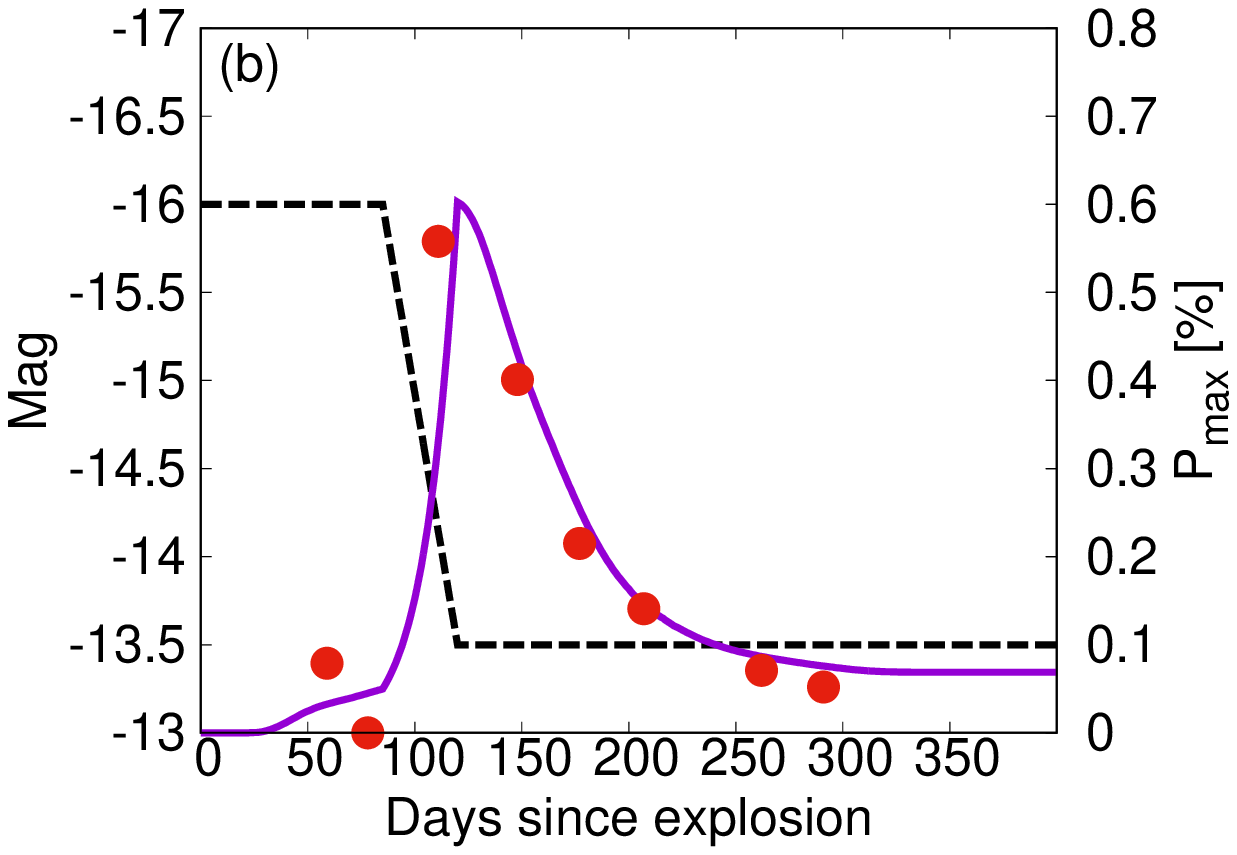}
\caption{(a) Time evolution of the observed polarization in SN 2004dj (the red points; Leonard et al. 2006), compared with those in the blob model for the best-fit parameters (the violet line). The dashed line shows the input light curve. (b) Same as (a), but for the bipolar CS dust model.}
\end{figure}

As for SN 2006ov, $P_{\mathrm{max}} \sim 1.6$ \% and $\Delta t \sim 50$ days are obtained from the observations in wavelengths around $\sim 7000-8000 \AA$ \citep{Chornock2010}. The luminosity drop of SN 2006ov in the $R$ band is $\sim 3.5$ mag \citep{Spiro2014}. This means that the relative flux of the scattered echo to the SN flux in the nebula phase, thus the polarization degree, is enhanced by a factor of $10^{3.5/2.5}/10^{2.5/2.5}\sim 2.5$ as compared to the case whose luminosity drop is $2.5$ mag. Figures 14 and 15 are the same as Figures 12 and 13, respetively, but for SN 2006ov, assuming the luminosity drop is $3.5$ mag in the models. It is also found that the values of $P_{\mathrm{max}}$ and $\Delta t$ in SN 2006ov can be reproduced using the blob model for $l_0 \sim 2.5 \times 10^{17}$ cm, $\tau_0 \sim 2.0$ and $\theta_{\mathrm{obs}} = 70$ degree, or the bipolar CS dust model for $r_{\mathrm{in}} \sim 1.5 \times 10^{17}$ cm, $\tau_0 \sim 3.0$ and $\theta_{\mathrm{obs}} = 70$ degree.

\begin{figure}[t]
  \includegraphics[scale=0.7]{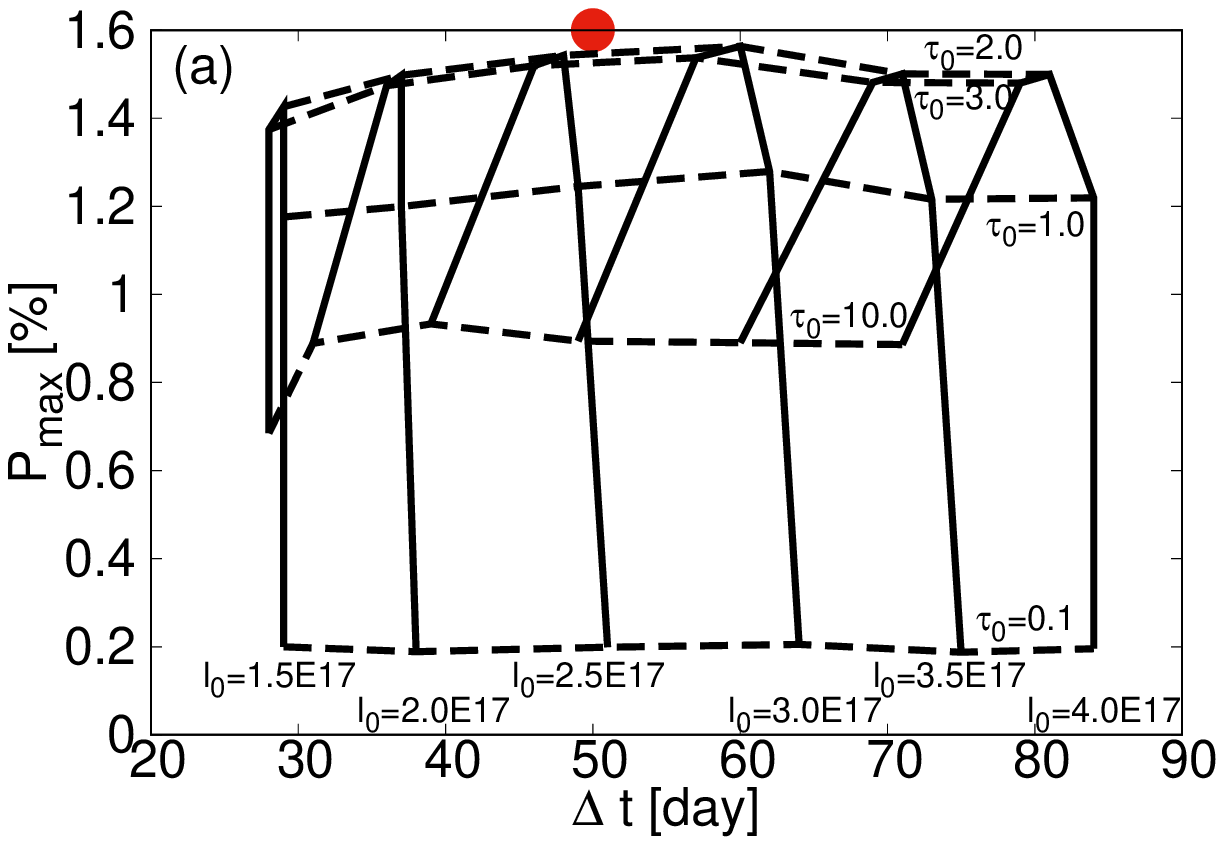}
  \includegraphics[scale=0.7]{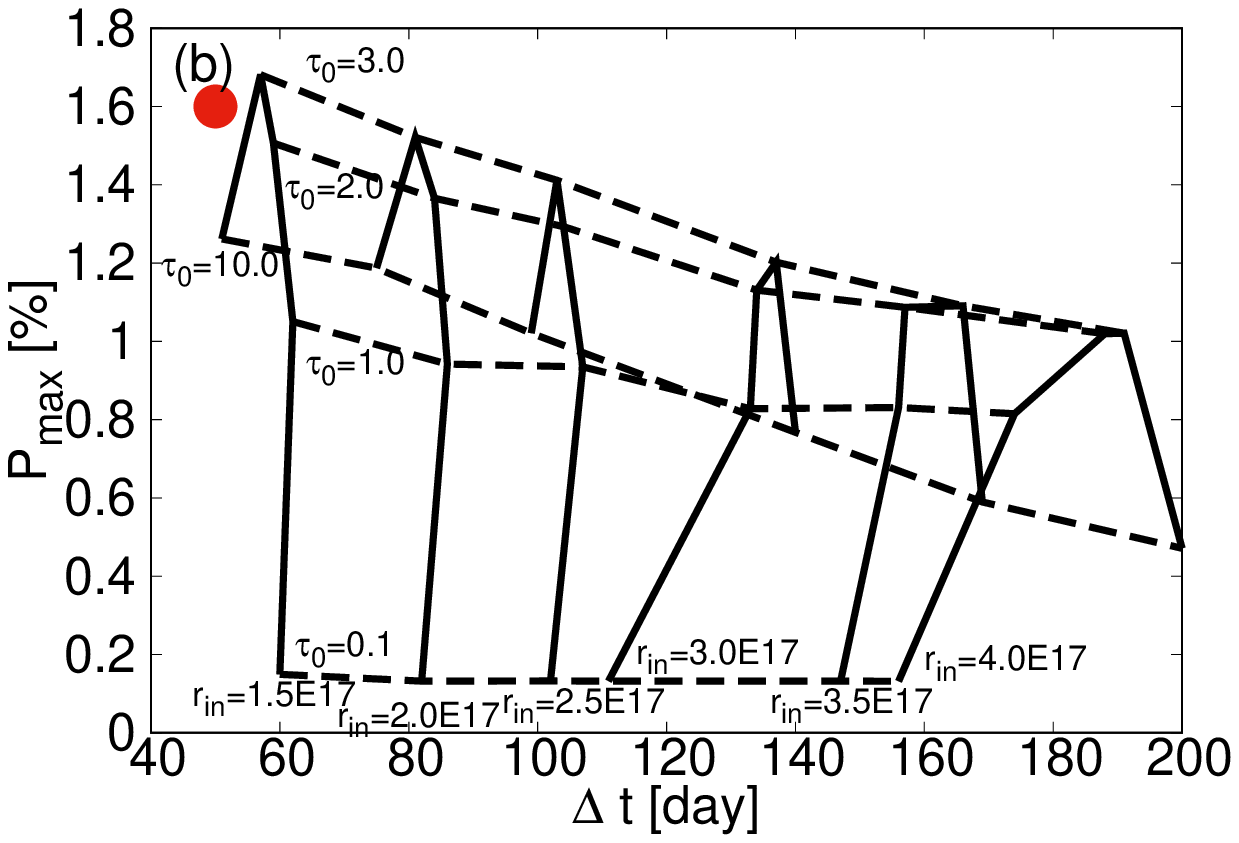}
\caption{Same as Fig. 12, but for SN 2006ov (Chornock et al. 2010).}
\end{figure}

\begin{figure}[t]
  \includegraphics[scale=0.7]{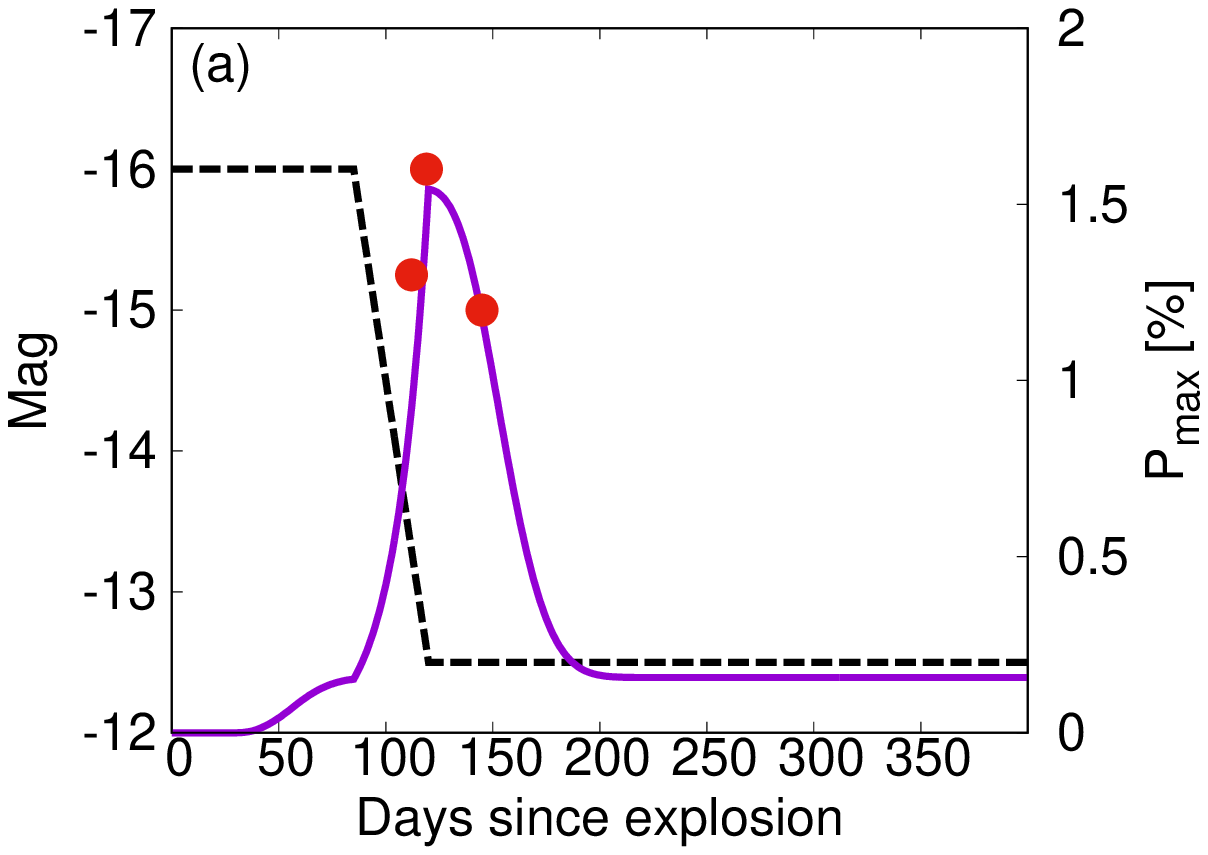}
  \includegraphics[scale=0.7]{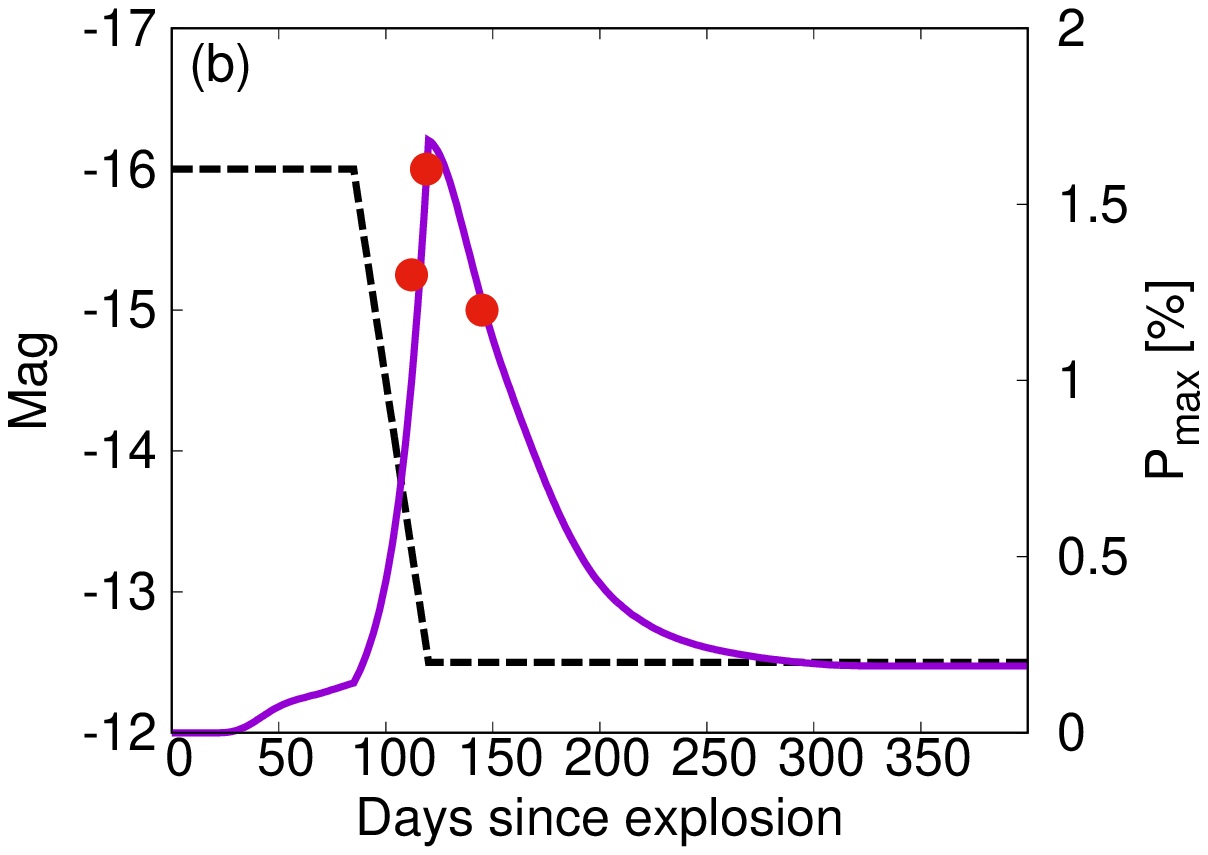}
\caption{Same as Fig. 13, but for SN 2006ov (Chornock et al. 2010).}
\end{figure}

While the best fit is fixed for $\theta_{\mathrm{obs}} = 70$ degree, the values of $P_{\mathrm{max}}$ do not sensitively depend on $\theta_{\mathrm{obs}}$ and are similar for $30 \lesssim \theta_{\mathrm{obs}} \lesssim 90$ degree (see \S 3.1), as long as the delay time of the scattered echo is longer than the timescale of the SN flux decrease (i.e., $35$ days). It is important to conduct polarimetric observations for SNe IIP in the late phase with sufficiently dense sampling to follow not only the increasing phase but also the decreasing phase.

\section{discussions}
We have investigated the effects of the polarized-scattered echoes from CS dust on polarization in SNe IIP, addressing dependences of the polarization on geometry and amount of CS dust. We have found that the time evolution of polarization in the well-observed SNe 2004dj and 2006ov could be reproduced by the dust scattering model. In the following subsections, discussions are given for expected IR emission from CS dust, further details of the dust scattering models, and future prospects.

\subsection{An infrared thermal echo from SN 2004dj}
If there is CS dust in the vicinity of an SN, IR emission by a thermal echo from the CS dust is also expected \citep[e.g.,][]{Dwek1985, Maeda2015, Nagao2017}. Given that infrared radiation from newly formed dust may also contribute to an observed IR luminosity, a predicted IR luminosity from CS dust derived from a polarization feature must be lower than the observed IR luminosity; Otherwise, this dust scattering model is rejected to explain the polarization feature.

The IR luminosity in SN 2004dj has been reported \citep[e.g.,][]{Kotak2005, Szalai2011, Meikle2011}. They have attributed the IR luminosity to thermal emission from newly formed dust in an ejecta or a cool dense shell, or an IR echo from interstellar dust. As already mentioned above, the expected IR luminosity from CS dust should be lower than the observed IR luminosity. We conduct radiative transfer calculations to check the consistency between the dust scattering model for the polarization of SN 2004dj and the IR observations \citep{Szalai2011}. We use a three-dimensional Monte Carlo radiative transfer code presented in \citet{Nagao2016}, using LMC dust model \citep[the LMC1 model in][]{Nagao2016} for simplicity. In this dust model, the values of $\omega$ and $g$ are $0.66$ and $0.47$, respectively, while we use $\omega=0.5$ and $g=0.6$ in \S 3. The differences of the values do not largely affect the following discussions. As an input light of an SN IIP, we adopt the blackbody spectra with a temperature of $6000$ K \citep[e.g.,][]{Bersten2009}. The flux of the input SN light is set so that the $V$-band light curve follows the light curve presented in \S 2.3.

Figure 16 shows the calculated IR light curves using the best-fit blob and bipolar CS dust models for the polarization. The observational points in the {\it spitzer}-IRAC bands are taken from \citet{Szalai2011}. Our best-fit models are consistent with the observed IR data, implying some additional sources for IR luminosity (e.g., newly formed dust). As for SN 2006ov, there are no publicly available IR data around the expected IR luminosity peak ($\sim 120$ days since the explosion). Our analysis shows the importance of a combined analysis of the polarization and IR properties to uncover and clarify the mechanism of the polarization evolution in SNe IIP.

\begin{figure*}[t]
  \includegraphics[scale=0.7]{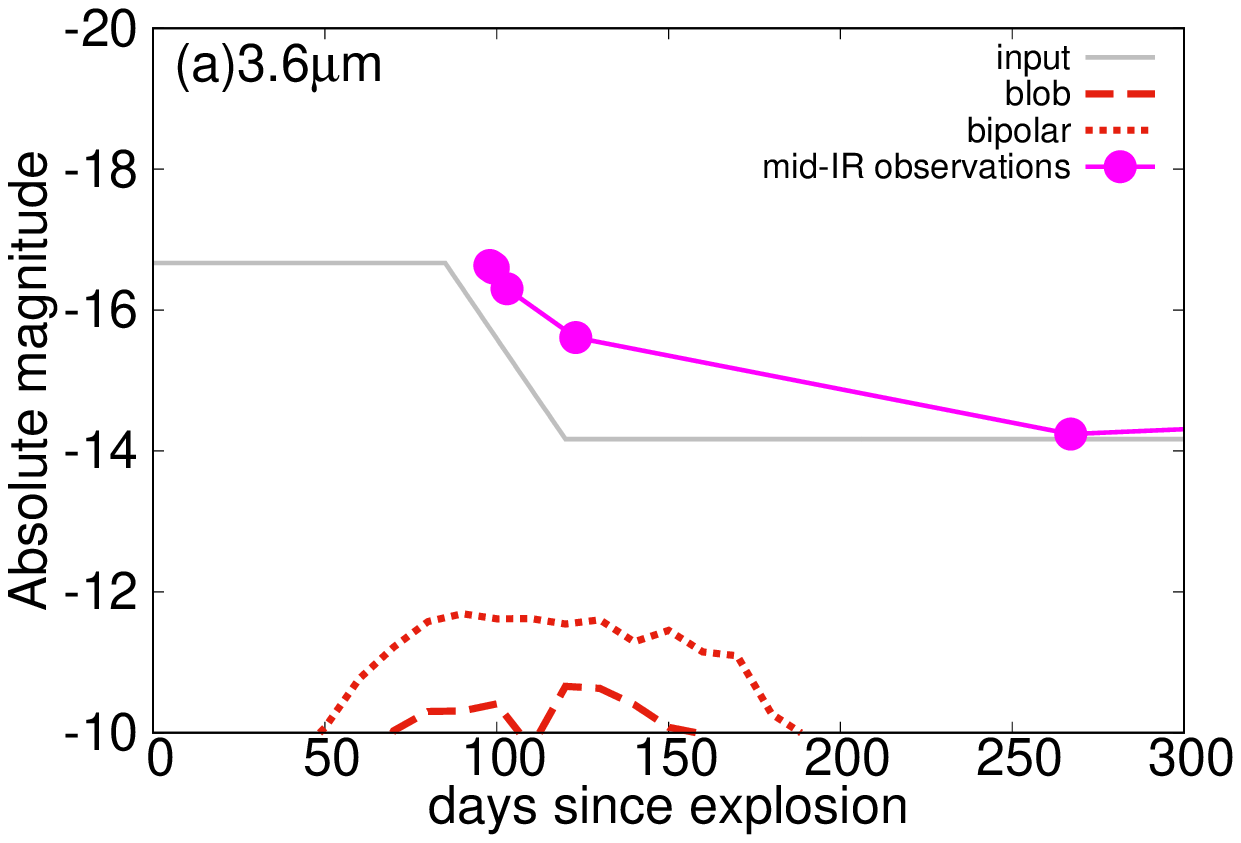}
  \includegraphics[scale=0.7]{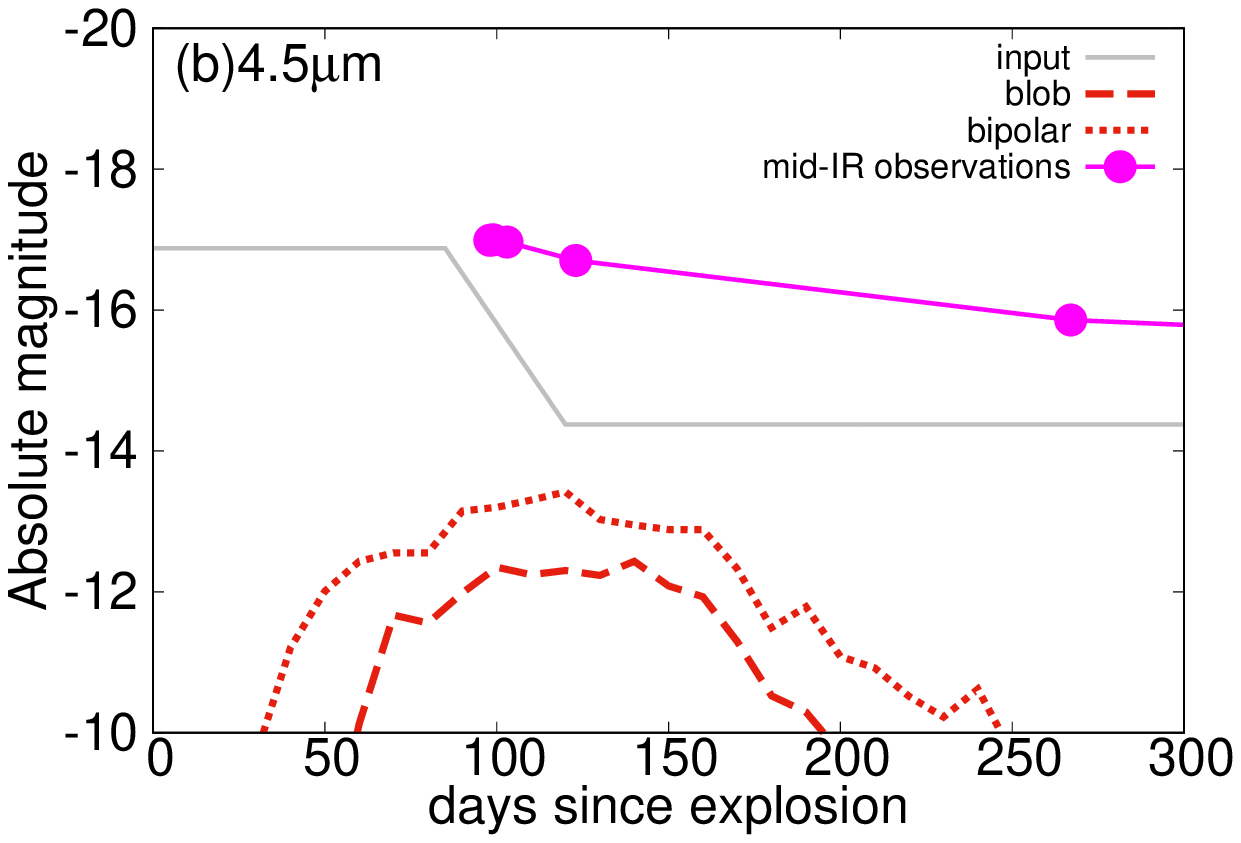}
  \includegraphics[scale=0.7]{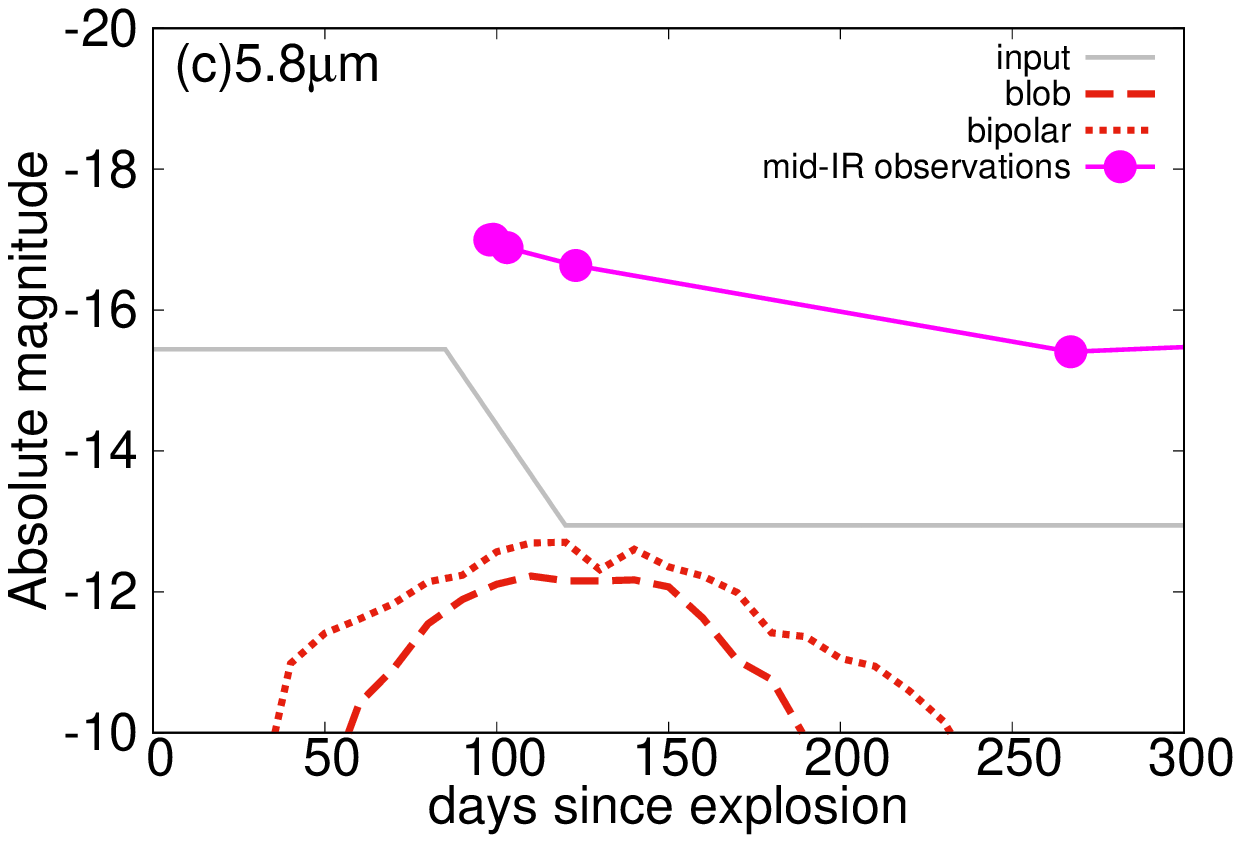}
  \includegraphics[scale=0.7]{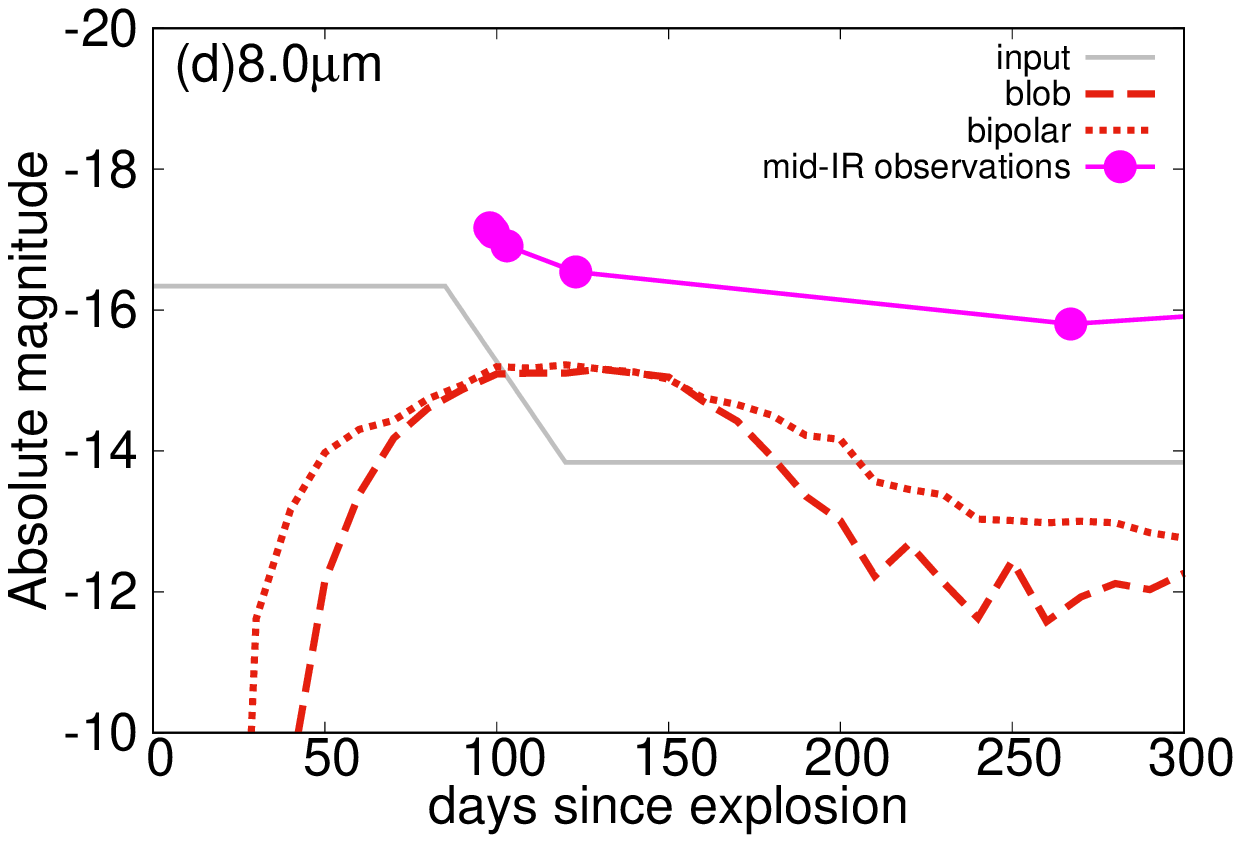}
\caption{The mid-IR echoes in the best-fit blob (red dashed lines) and bipolar CS dust (red doted lines) models for SN 2004dj, with the observed data (Vega magnitude; magenta points) reported by \citet{Szalai2011}.}
\end{figure*}

\subsection{The minimum radius of CS dust and dust evaporation by SN initial flash}
Since CS dust grains within a certain distance from the SN are expected to be destroyed by the initial UV flash after shock breakout, we cannot use arbitrary values for $l_0$ (in the blob model) or $r_{in}$ (in the disk and bipolar CS dust models). The evaporation radius of CS dust is a radius below which radiative-equilibrium temperature of CS dust by the initial flash (which depends on grain size of CS dust and luminosity of the flash) is higher than the evaporation temperature (which depends on grain size of CS dust and duration of the initial flash). In fact, our knowledge on this process is still too limited to precisely determine the evaporation radius. \citet{Wang1996} calculated the evaporation radius of CS dust for various initial flash models and showed that the evaporation radius of graphite dust is $\sim 1.0 \times 10^{17}$ cm ($\sim 0.03$ pc) for the typical flash models for SNe IIP. \citet{Fischera2002} also calculated the radius using similar equations and reached to a similar conclusion. 

Moreover, there are also observational evidences for pre-existing dust around some SNe, at $\sim 3 \times 10^{17}$ cm ($\sim 0.1$ pc) from the SN. For SN 1987A, a dusty ring is detected at $\sim 6 \times 10^{17}$ cm ($\sim 0.2$ pc) \citep[e.g.,][]{Sonneborn1998, Lawrence2000}. For SN IIP 2002hh, \citet{Barlow2005} reported thermal emission from CS dust, as inferred from mid-infrared (mid-IR) data taken by the Spitzer Telescope (SST). The minimum distance was derived to be $\sim 9 \times 10^{16}$ cm ($\sim 0.03$ pc) from the SN. For SN 2008S which is known to have been surrounded by a dusty environment before the explosion \citep[e.g.,][]{Prieto2008}, \citet{Wesson2010} claimed that a inner radius of CS dust moved from $\sim 1.3 \times 10^{15}$ cm ($\sim 0.0004$ pc; pre-SN) to $\sim 1.9 \times 10^{16}$ cm ($\sim 0.006$ pc; post-SN) by  the grain evaporation caused by the SN flash, on the basis of mid-IR observation by SST. The inner radius of CS dust around Type IIn SN 2010jl are also estimated to be $\sim 6 \times 10^{17}$ cm ($\sim 0.2$ pc) using SST data \citep{Andrews2011}. These values are consistent with our setup for the inner radius of CS dust, although there are some uncertainties for the properties of the flash (luminosity, duration and temperature of the flash) and the dust grain (mass absorption coefficient and radius of CS dust grains).  This issue will be further investigated elsewhere (Maeda et al. in prep.).

\subsection{Considerations on model assumptions}
As our reference models, we use $g=0.6$ as a scattering asymmetry parameter. This is a typical value in optical bands, which is the wavelength range of interest in this study. In fact, the value of $g$ is observationally not well known. We also investigate a case with $g=0.3$, which means almost isotropic scattering. In this case, the polarization degree becoms largest for $\theta_{\mathrm{obs}} = 90$ degree. Figure 17 shows the same quantities as Fig. 6, but with $g=0.3$ and $\theta_{\mathrm{obs}} = 90$ degree. The larger $P_{\mathrm{max}}$ is produced, reaching to $\sim 0.9$ \%. As for the value of albedo, we use $\omega=0.5$ as a typical value. However, the value is also observationally uncertain \citep[e.g.,][for a review]{Draine2011}. If we adopt a larger value of albedo, $P_{\rm{max}}$ and $\Delta t$ simply become slightly higher and longer, respectively.

\begin{figure}[t]
  \includegraphics[scale=0.7]{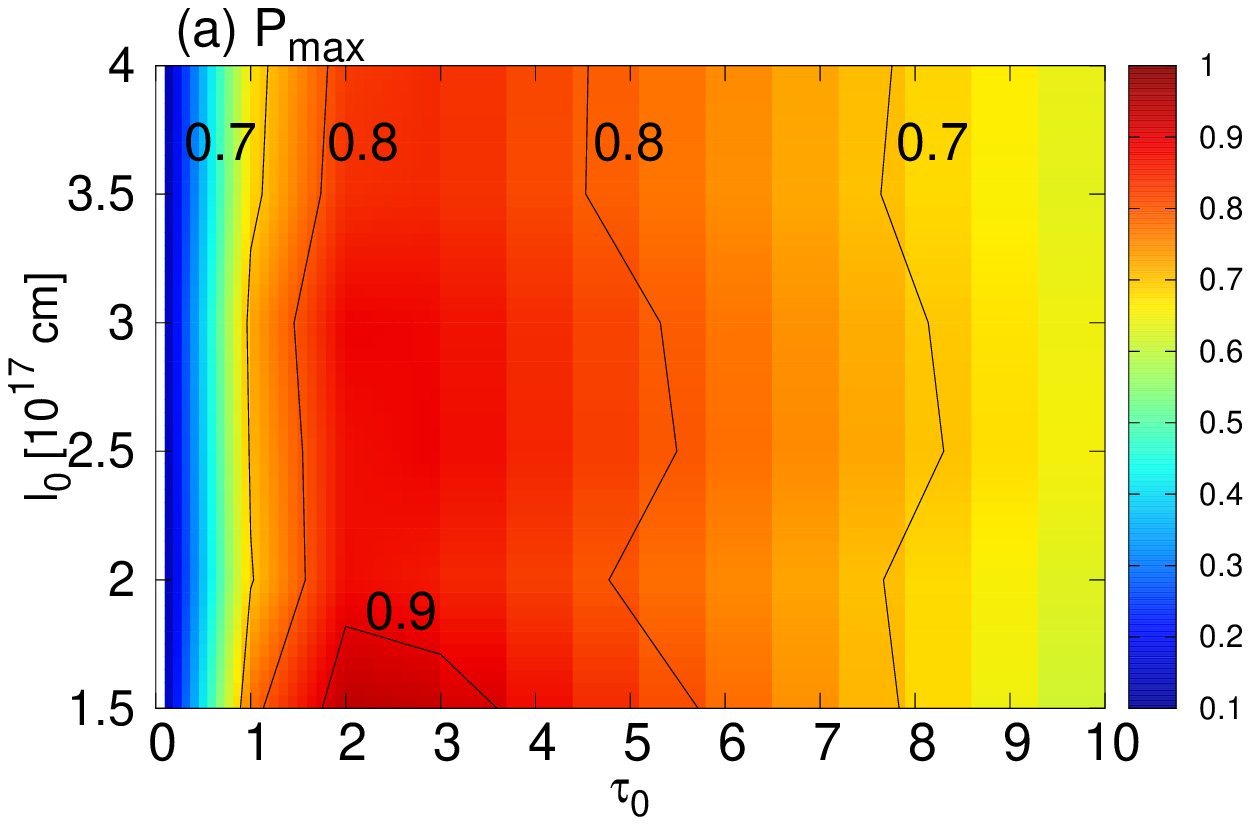}
  \includegraphics[scale=0.7]{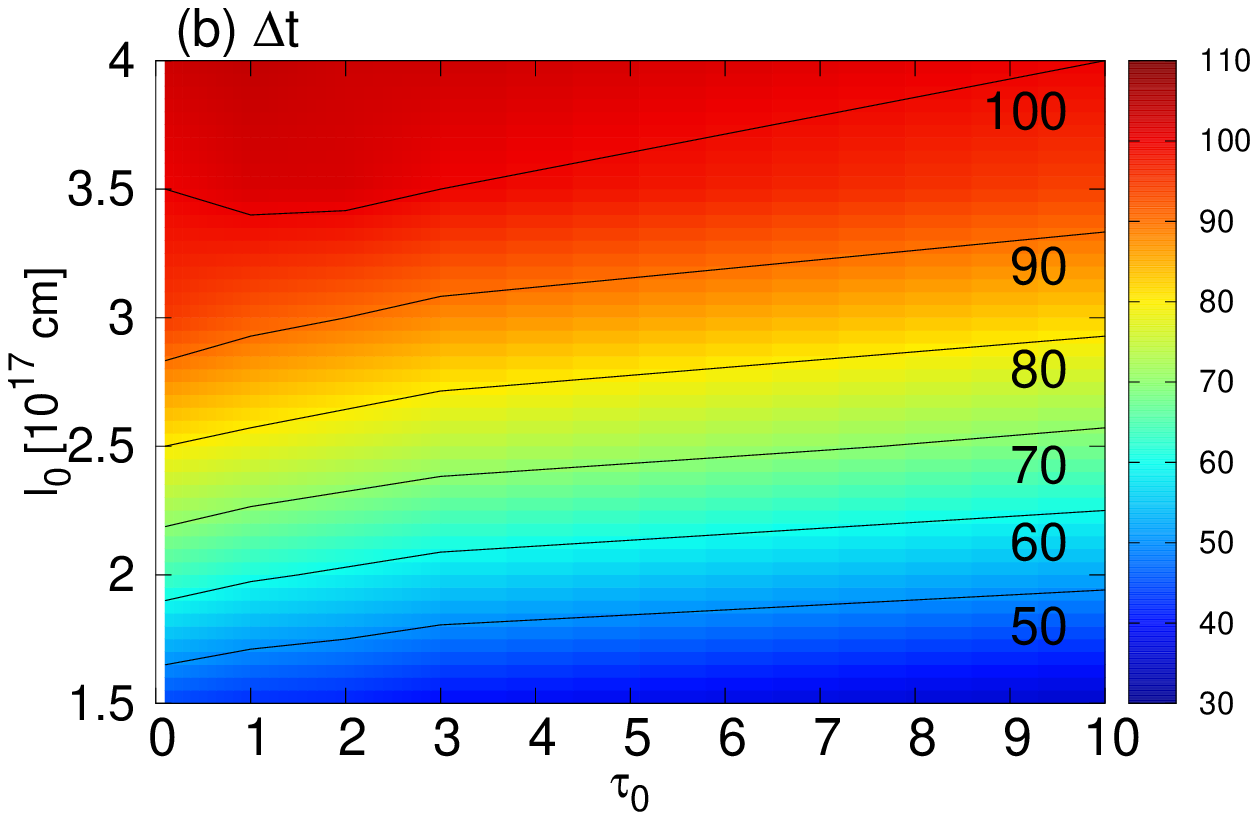}
\caption{Same as Fig. 6, but for the blob models with $g=0.3$.}
\end{figure}

In the present study, we assume one blob (disk, bipolar CS dust) whose solid angle covering the SN light is $4\pi \times 0.01$ to represent a CS dust distribution. If the solid angle is bigger, it may be expected that the polarization degree is bigger. However, the counter effect is the increasing importance of multiple scattering within the blob, which reduces the polarization degree. Moreover, if there are several blobs randomly distributed around an SN, an expected polarization curve is the sum of the polarized echoes from the blobs, in general leading to cancellation of polarization. We use a fixed value for $r_{\rm{out}} - r_{\rm{in}} = 3.0 \times 10^{17}$ cm in the disk and bipolar CS dust models. This value does not sensitively affect the results in \S 3, because optical depth of each part in the disk (the bipolar CS dust) is inversely proportional to a radius $r$ and therefore the echo signal is dominated by the CS dust around $r_{\rm{in}}$.

For simplicity, the luminosity of the input SN light is assumed to be constant during the plateau and nebula phases. In fact, the luminosity tends to slightly decrease during each phase, where the decay rate of the light curve is different for different SNe. Since an SN usually decays faster in the nebula phase than in the plateau phase, the value of $P_{\mathrm{max}}$ might be even larger than that found in our results.

\subsection{Prospects for the future}
It is important to study time evolution of polarization in a wide range of wavelength for the same object, for revealing the dominant mechanism to create polarization. Our results can apply to any optical bands, taking into account effects of different wavelengths for dust parameters and light curve shapes: the optical depth of CS dust ($\tau_0(\nu)$), the values of albedo and scattering asymmetry parameter ($\omega(\nu)$, $g(\nu)$) and depth of the light curve drop ($\Delta M_{\mathrm{drop}}(\nu)$). However, it is not so simple to predict strengths of maximum polarization degree in different wavelengths, because polarization degree depends differently on the above parameters ($\tau_0(\nu)$,$\omega(\nu)$, $g(\nu)$, $\Delta M_{\mathrm{drop}}(\nu)$). The larger optical depth, albedo and luminosity drop (all generally expected in a bluer band) should lead to higher polarization degree. At the same time, forward scattering become more important in a bluer band, which would have the opposite effect to reduce the polarization degree. Therefore, prediction of the wavelength dependence of the polarization in SNe IIP requires good knowledge on properties of CS dust. Alternatively, we can compute the wavelength-dependent polarization signals for given dust models, which we postpone to future. 

The applicability of the present study is not limited to SNe IIP. For light sources within asymmetrically-distributed  CS dust, especially for light sources whose luminosities decrease as a function of time, the polarized-scattered echo is necessarily produced. For example, polarization in superlumionous SNe that show interaction features with dense CS medium (i.e., SLSNe-II) could be used to constrain the spacial distribution of the CS dust. Even if this process would not have a dominant role in the observed polarization signals, this effect should in principle contribute to it, the strength of which depends on the nature of CS dust. Therefore, this effect must be taken into account in discussing a multi-dimensional structure of an SN explosion through polarimetric observations.

\section{Conclusions}
We have investigated the effects of the scattered echoes from CS dust on the polarization of SNe IIP through radiation transfer simulations for various geometry and amount of CS dust. It has been found that asymmetrically-distributed CS dust, which is generally inferred for RSGs, could reproduce the observed polarization features. 

We have applied our results to SNe 2004dj and 2006ov, deriving the geometry and amount of CS dust to explain their observed polarization features. The polarization feature in SN 2004dj could be reproduced using the blob model for $l_0 \sim 3.0 \times 10^{17}$ cm, $\tau_0 \sim 2.0$ and $\theta_{\mathrm{obs}} = 70$ degree, or the bipolar CS dust model for $r_{\mathrm{in}} \sim 1.5 \times 10^{17}$ cm, $\tau_0 \sim 2.0$ and $\theta_{\mathrm{obs}} = 70$ degree. The corresponding dust mass $M_{\mathrm{dust}}$ (and mass-loss rate $\dot{M_{\mathrm{gas}}}$) in the blob and bipolar CS dust models are $\sim 7.5 \times 10^{-4}$ M$_{\odot}$ ($\sim 2.0 \times10^{-5}$ M$_{\odot}$ yr$^{-1}$) and $\sim 8.5 \times 10^{-4}$ M$_{\odot}$ ($\sim 8.9 \times10^{-6}$ M$_{\odot}$ yr$^{-1}$), respectively (see \S 2.2). The polarization feature in SN 2006ov can be also reproduced using the blob model for $l_0 \sim 2.5 \times 10^{17}$ cm, $\tau_0 \sim 2.0$ and $\theta_{\mathrm{obs}} = 70$ degree, or the bipolar CS dust model for $r_{\mathrm{in}} \sim 1.5 \times 10^{17}$ cm, $\tau_0 \sim 3.0$ and $\theta_{\mathrm{obs}} = 70$ degree. The corresponding dust mass $M_{\mathrm{dust}}$ (and mass-loss rate $\dot{M_{\mathrm{gas}}}$) in the blob and bipolar CS dust models are $\sim 5.2 \times 10^{-4}$ M$_{\odot}$ ($\sim 1.7 \times10^{-5}$ M$_{\odot}$ yr$^{-1}$) and $\sim 1.3 \times 10^{-3}$ M$_{\odot}$ ($\sim 1.3 \times10^{-5}$ M$_{\odot}$ yr$^{-1}$), respectively. Here, the value of $\theta_{\mathrm{obs}}$ is not so strictly limited, because the values of $P_{\mathrm{max}}$ do not sensitively depend on $\theta_{\mathrm{obs}}$ for $30 \lesssim \theta_{\mathrm{obs}} \lesssim 90$ degree (see \S 3.1), as long as the delay time of the scattered echo is longer than the timescale of the SN flux decrease (i.e., $35$ days).

We have also clarified that the dust scattering model for the polarization of SN 2004dj is consistent with the IR observations. It is important to obtain not only polarimetric information but also IR information for SNe IIP, to understand the mechanism of the polarization evolution.

\acknowledgments
The authors thank Lifan Wang for stimulating discussion to initiate this project and an anonymous referee for constructive comments. The authors also thank Llu\'is Galbany for organizing the workshop ``Supernova is In Da House", where this work was initiated. The participation of TN and KM to this workshop was supported by JSPS Open Partnership Bilateral Joint Research Project between Japan and Chile (K.M.). The authors thank the Yukawa Institute for Theoretical Physics at Kyoto University. Discussions during the YITP workshop YITP-T-16-05 on ``Transient Universe in the Big Survey Era: Understanding the Nature of Astrophysical Explosive Phenomena" were useful to complete this work. The authors thank Keiichi Ohnaka for his useful comments on the geometry of CS dust around RSGs. TN thanks Takashi Kozasa and Takaya Nozawa for giving a lot of advices on the Monte Carlo code to calculate polarization by dust scattering. The simulations were in part carried out on the PC cluster at the Center for Computational Astrophysics, National Astronomical Observatory of Japan. This research has made use of the Spanish Virtual Observatory (http://svo.cab.inta-csic.es) supported from the Spanish MINECO through grant AyA2014-55216 for the filter profiles. The work has been supported by Japan Society for the Promotion of Science (JSPS) KAKENHI Grant 17J06373 (T.N.), 26800100 and 17H02864 (K.M.), 15H02075 (M.T.), MEXT KAKENHI Grant 15H00788 (M.T.), and Inoue Foundation for Science (M.T.).



\begin{thebibliography}{}
\bibitem[Anderson et al.(2014)]{Anderson2014} Anderson, J.~P., Gonz{\'a}lez-Gait{\'a}n, S., Hamuy, M., et al.\ 2014, \apj, 786, 67
\bibitem[Andrews et al.(2011)]{Andrews2011} Andrews, J.~E., Clayton, G.~C., Wesson, R., et al.\ 2011, \aj, 142, 45
\bibitem[Barlow et al.(2005)]{Barlow2005} Barlow, M.~J., Sugerman, B.~E.~K., Fabbri, J., et al.\ 2005, \apjl, 627, L113
\bibitem[Bersten \& Hamuy(2009)]{Bersten2009} Bersten, M.~C., \& Hamuy, M.\ 2009, \apj, 701, 200
\bibitem[Bruenn et al.(2013)]{Bruenn2013} Bruenn, S.~W., Mezzacappa, A., Hix, W.~R., et al.\ 2013, \apjl, 767, L6
\bibitem[Bulla et al.(2015)]{Bulla2015} Bulla, M., Sim, S.~A., \& Kromer, M.\ 2015, \mnras, 450, 967  
\bibitem[Buras et al.(2006)]{Buras2006} Buras, R., Rampp, M., Janka, H.-T., \& Kifonidis, K.\ 2006, \aap, 447, 1049
\bibitem[Chandrasekhar(1960)]{Chandrasekhar1960} Chandrasekhar, S.\ 1960, Radiative Tranfer (New York: Dover)  
\bibitem[Chornock et al.(2010)]{Chornock2010} Chornock, R., Filippenko, A.~V., Li, W., \& Silverman, J.~M.\ 2010, \apj, 713, 1363
\bibitem[Code \& Whitney(1995)]{Code1995} Code, A.~D., \& Whitney, B.~A.\ 1995, \apj, 441, 400
\bibitem[Couch \& O'Connor(2014)]{Couch2014} Couch, S.~M., \& O'Connor, E.~P.\ 2014, \apj, 785, 123
\bibitem[Cruzalebes et al.(1998)]{Cruzalebes1998} Cruzalebes, P., Lopez, B., Bester, M., Gendron, E., \& Sams, B.\ 1998, \aap, 338, 132
\bibitem[Dessart \& Hillier(2011)]{Dessart2011} Dessart, L., \& Hillier, D.~J.\ 2011, \mnras, 415, 3497
\bibitem[Draine(2011)]{Draine2011} Draine, B.~T.\ 2011, Physics of the Interstellar and Intergalactic Medium by Bruce T.~Draine.~Princeton University Press, 2011.~ISBN: 978-0-691-12214-4, 
\bibitem[Dwek(1985)]{Dwek1985} Dwek, E.\ 1985, \apj, 297, 719
\bibitem[Fischera et al.(2002)]{Fischera2002} Fischera, J., Tuffs, R.~J., \& V{\"o}lk, H.~J.\ 2002, \aap, 395, 189
\bibitem[Hanke et al.(2013)]{Hanke2013} Hanke, F., M{\"u}ller, B., Wongwathanarat, A., Marek, A., \& Janka, H.-T.\ 2013, \apj, 770, 66
\bibitem[Hinz et al.(1998)]{Hinz1998} Hinz, P.~M., Angel, J.~R.~P., Hoffmann, W.~F., et al.\ 1998, \nat, 395, 251
\bibitem[H{\"o}flich(1991)]{Hoflich1991} H{\"o}flich, P.\ 1991, \aap, 246, 481
\bibitem[H{\"o}flich et al.(1996)]{Hoflich1996} H{\"o}flich, P., Wheeler, J.~C., Hines, D.~C., \& Trammell, S.~R.\ 1996, \apj, 459, 307
\bibitem[Humphreys et al.(2007)]{Humphreys2007} Humphreys, R.~M., Helton, L.~A., \& Jones, T.~J.\ 2007, \aj, 133, 2716
\bibitem[Kasen et al.(2006)]{Kasen2006} Kasen, D., Thomas, R.~C., \& Nugent, P.\ 2006, \apj, 651, 366
\bibitem[Kastner \& Weintraub(1998)]{Kastner1998} Kastner, J.~H., \& Weintraub, D.~A.\ 1998, \aj, 115, 1592 
\bibitem[Kervella et al.(2009)]{Kervella2009} Kervella, P., Verhoelst, T., Ridgway, S.~T., et al.\ 2009, \aap, 504, 115
\bibitem[Kervella et al.(2011)]{Kervella2011} Kervella, P., Perrin, G., Chiavassa, A., et al.\ 2011, \aap, 531, A117
\bibitem[Kervella et al.(2016)]{Kervella2016} Kervella, P., Lagadec, E., Montarg{\`e}s, M., et al.\ 2016, \aap, 585, A28
\bibitem[Kotak et al.(2005)]{Kotak2005} Kotak, R., Meikle, P., van Dyk, S.~D., H{\"o}flich, P.~A., \& Mattila, S.\ 2005, \apjl, 628, L123
\bibitem[Kumar et al.(2016)]{Kumar2016} Kumar, B., Pandey, S.~B., Eswaraiah, C., \& Kawabata, K.~S.\ 2016, \mnras, 456, 3157
\bibitem[Lawrence et al.(2000)]{Lawrence2000} Lawrence, S.~S., Sugerman, B.~E., Bouchet, P., et al.\ 2000, \apjl, 537, L123
\bibitem[Lentz et al.(2015)]{Lentz2015} Lentz, E.~J., Bruenn, S.~W., Hix, W.~R., et al.\ 2015, \apjl, 807, L31
\bibitem[Leonard et al.(2006)]{Leonard2006} Leonard, D.~C., Filippenko, A.~V., Ganeshalingam, M., et al.\ 2006, \nat, 440, 505
\bibitem[Li et al.(2011)]{Li2011} Li, W., Leaman, J., Chornock, R., et al.\ 2011, \mnras, 412, 1441
\bibitem[Liebend{\"o}rfer et al.(2001)]{Liebendorfer2001} Liebend{\"o}rfer, M., Mezzacappa, A., Thielemann, F.-K., et al.\ 2001, \prd, 63, 103004
\bibitem[Maeda et al.(2008)]{Maeda2008} Maeda, K., Kawabata, K., Mazzali, P.~A., et al.\ 2008, Science, 319, 1220 
\bibitem[Maeda et al.(2015)]{Maeda2015} Maeda, K., Nozawa, T., Nagao, T., \& Motohara, K.\ 2015, \mnras, 452, 3281
\bibitem[Marek \& Janka(2009)]{Marek2009} Marek, A., \& Janka, H.-T.\ 2009, \apj, 694, 664
\bibitem[Marsh et al.(2001)]{Marsh2001} Marsh, K.~A., Bloemhof, E.~E., Koerner, D.~W., \& Ressler, M.~E.\ 2001, \apj, 548, 861
\bibitem[Marshall et al.(2004)]{Marshall2004} Marshall, J.~R., van Loon, J.~T., Matsuura, M., et al.\ 2004, \mnras, 355, 1348
\bibitem[Mauerhan et al.(2017)]{Mauerhan2017} Mauerhan, J.~C., Van Dyk, S.~D., Johansson, J., et al.\ 2017, \apj, 834, 118
\bibitem[Mauron \& Josselin(2011)]{Mauron2011} Mauron, N., \& Josselin, E.\ 2011, \aap, 526, A156
\bibitem[Meikle et al.(2011)]{Meikle2011} Meikle, W.~P.~S., Kotak, R., Farrah, D., et al.\ 2011, \apj, 732, 109
\bibitem[Melson et al.(2015)]{Melson2015} Melson, T., Janka, H.-T., \& Marek, A.\ 2015, \apjl, 801, L24
\bibitem[Monnier et al.(2004)]{Monnier2004} Monnier, J.~D., Millan-Gabet, R., Tuthill, P.~G., et al.\ 2004, \apj, 605, 436
\bibitem[M{\"u}ller et al.(2012)]{Muller2012} M{\"u}ller, B., Janka, H.-T., \& Heger, A.\ 2012, \apj, 761, 72  
\bibitem[M{\"u}ller(2016)]{Muller2016} M{\"u}ller, B.\ 2016, \pasa, 33, e048
\bibitem[Nagao et al.(2016)]{Nagao2016} Nagao, T., Maeda, K., \& Nozawa, T.\ 2016, \apj, 823, 104
\bibitem[Nagao et al.(2017)]{Nagao2017} Nagao, T., Maeda, K., \& Yamanaka, M.\ 2017, \apj, 835, 143
\bibitem[Ohnaka et al.(2009)]{Ohnaka2009} Ohnaka, K., Hofmann, K.-H., Benisty, M., et al.\ 2009, \aap, 503, 183
\bibitem[Ohnaka et al.(2011)]{Ohnaka2011} Ohnaka, K., Weigelt, G., Millour, F., et al.\ 2011, \aap, 529, A163
\bibitem[Ohnaka et al.(2013)]{Ohnaka2013} Ohnaka, K., Hofmann, K.-H., Schertl, D., et al.\ 2013, \aap, 555, A24  
\bibitem[Ohnaka(2014)]{Ohnaka2014} Ohnaka, K.\ 2014, \aap, 568, A17
\bibitem[Patat(2005)]{Patat2005} Patat, F.\ 2005, \mnras, 357, 1161
\bibitem[Prieto et al.(2008)]{Prieto2008} Prieto, J.~L., Kistler, M.~D., Thompson, T.~A., et al.\ 2008, \apjl, 681, L9
\bibitem[Rampp \& Janka(2000)]{Rampp2000} Rampp, M., \& Janka, H.-T.\ 2000, \apjl, 539, L33
\bibitem[Roberts et al.(2016)]{Roberts2016} Roberts, L.~F., Ott, C.~D., Haas, R., et al.\ 2016, \apj, 831, 98
\bibitem[Sanders et al.(2015)]{Sanders2015} Sanders, N.~E., Soderberg, A.~M., Gezari, S., et al.\ 2015, \apj, 799, 208
\bibitem[Shapiro \& Sutherland(1982)]{Shapiro1982} Shapiro, P.~R., \& Sutherland, P.~G.\ 1982, \apj, 263, 902
\bibitem[Smith et al.(2001)]{Smith2001} Smith, N., Humphreys, R.~M., Davidson, K., et al.\ 2001, \aj, 121, 1111
\bibitem[Sonneborn et al.(1998)]{Sonneborn1998} Sonneborn, G., Pun, C.~S.~J., Kimble, R.~A., et al.\ 1998, \apjl, 492, L139
\bibitem[Spiro et al.(2014)]{Spiro2014} Spiro, S., Pastorello, A., Pumo, M.~L., et al.\ 2014, \mnras, 439, 2873  
\bibitem[Sumiyoshi et al.(2005)]{Sumiyoshi2005} Sumiyoshi, K., Yamada, S., Suzuki, H., et al.\ 2005, \apj, 629, 922
\bibitem[Suwa et al.(2010)]{Suwa2010} Suwa, Y., Kotake, K., Takiwaki, T., et al.\ 2010, \pasj, 62, L49
\bibitem[Szalai et al.(2011)]{Szalai2011} Szalai, T., Vink{\'o}, J., Balog, Z., et al.\ 2011, \aap, 527, A61
\bibitem[Takiwaki et al.(2012)]{Takiwaki2012} Takiwaki, T., Kotake, K., \& Suwa, Y.\ 2012, \apj, 749, 98
\bibitem[Takiwaki et al.(2014)]{Takiwaki2014} Takiwaki, T., Kotake, K., \& Suwa, Y.\ 2014, \apj, 786, 83
\bibitem[Tanaka et al.(2017)]{Tanaka2017} Tanaka, M., Maeda, K., Mazzali, P.~A., Kawabata, K.~S., \& Nomoto, K.\ 2017, \apj, 837, 105
\bibitem[Thompson et al.(2003)]{Thompson2003} Thompson, T.~A., Burrows, A., \& Pinto, P.~A.\ 2003, \apj, 592, 434
\bibitem[van de Hulst(1957)]{van de Hulst1957} van de Hulst, H.~C.\ 1957, Light Scattering by Small Particles (New York: John Wiley \& Sons)
\bibitem[Wang \& Wheeler(1996)]{Wang1996} Wang, L., \& Wheeler, J.~C.\ 1996, \apjl, 462, L27
\bibitem[Wang \& Wheeler(2008)]{Wang2008} Wang, L., \& Wheeler, J.~C.\ 2008, \araa, 46, 433
\bibitem[Wesson et al.(2010)]{Wesson2010} Wesson, R., Barlow, M.~J., Ercolano, B., et al.\ 2010, \mnras, 403, 474
\bibitem[White(1979)]{White1979} White, R.~L.\ 1979, \apj, 229, 954
\bibitem[Wittkowski et al.(1998)]{Wittkowski1998} Wittkowski, M., Langer, N., \& Weigelt, G.\ 1998, \aap, 340, L39 
  
\end{thebibliography}
\end{document}